\documentclass[preprint]{aastex}
\usepackage{natbib}
\slugcomment{draft version  \today}
\usepackage[section] {placeins}
\begin{document}
\title{Planet Hunters: Assessing the \emph{Kepler} Inventory of Short Period Planets  }
\author{Megan E. Schwamb\altaffilmark{1,}\altaffilmark{2,3},Chris J. Lintott\altaffilmark{4,}\altaffilmark{5}, Debra A. Fischer\altaffilmark{6}, Matthew J. Giguere\altaffilmark{6}, Stuart Lynn\altaffilmark{5,}\altaffilmark{4},  Arfon M. Smith\altaffilmark{5,}\altaffilmark{4}, John M. Brewer\altaffilmark{6}, Michael Parrish\altaffilmark{5}, Kevin Schawinski\altaffilmark{2,}\altaffilmark{3,7}, and Robert J. Simpson\altaffilmark{4} }
\altaffiltext{1}{Yale Center for Astronomy and Astrophysics, Yale University,P.O. Box 208121, New Haven, CT 06520}
\altaffiltext{2}{Department of Physics, Yale University, New Haven, CT 06511}
\altaffiltext{3} {NSF Astronomy and Astrophysics Postdoctoral Fellow}
\altaffiltext{4}{Oxford Astrophysics, Denys Wilkinson Building, Keble Road, Oxford OX1 3RH}
\altaffiltext{5}{Adler Planetarium, 1300 S. Lake Shore Drive, Chicago, IL 60605}
\altaffiltext{6}{Department of Astronomy, Yale University, New Haven, CT 06511}
\altaffiltext{7}{Einstein Fellow}
\email{megan.schwamb@yale.edu}

\begin{abstract} 
We present the results from a search of data from the first  33.5  days of the \emph{Kepler} science mission (Quarter 1) for exoplanet transits by the Planet Hunters citizen science project. Planet Hunters enlists  members of the general public to visually identify transits in the publicly released \emph{Kepler} light curves via the World Wide Web. Over 24,000 volunteers reviewed the \emph{Kepler} Quarter 1 data set.  We examine the abundance of $\ge$ 2  R$_{\oplus}$ planets on short period ($<$ 15 days) orbits based on Planet Hunters detections. We present these results along with an analysis of the detection efficiency of human classifiers to identify planetary transits including a comparison to the \emph{Kepler} inventory of planet candidates. Although performance drops rapidly for smaller radii, $\ge$ 4 R$_{\oplus}$  Planet Hunters $\ge$ 85$\%$ efficient  at identifying transit signals for planets with periods less than 15 days for the \emph{Kepler} sample of target stars. Our high efficiency rate for simulated transits along with recovery of the majority of \emph{Kepler}  $\ge$4 R$_{\oplus}$ planets suggest suggests the \emph{Kepler} inventory of  $\ge$4 R$_{\oplus}$ short period planets is nearly complete.  
  
\end{abstract}
\section{Introduction}

In the past nearly two decades, there has been an explosion in the number of known planets orbiting stars beyond our own solar system, with over 700 extrasolar planets (exoplanets)  known to date \citep{2011epsc.conf....3S,2011PASP..123..412W} .  While most early discoveries were made using the radial velocity method (\citeauthor{2010exop.book...27L} \citeyear{2010exop.book...27L}  and references therein), the advent of large digital CCD-cameras has enabled astronomers to make effective use of the transit method, recording the dimming in light that is seen if a planet passes in front of its parent star as viewed from Earth.  Ground-based transit surveys (\citeauthor{2010arXiv1001.2010W} \citeyear{2010arXiv1001.2010W} and references therein) have successfully detected the $\sim$1$\%$ drop in flux due to Jupiter and Neptune-sized planets, but we have entered a new era of exoplanet discovery with the spaced-based  \emph{CoRoT}  \citep{2003AdSpR..31..345B}  and \emph{Kepler} \citep{2010ApJ...713L..79K,2010Sci...327..977B} telescopes surveying tens to hundreds of thousands of main sequence stars with photometric precision sufficient to observe the 0.0084$\%$  transit depth of an Earth-sized planet transiting a Sun-like star  \citep[e.g.][]{2011arXiv1112.4550F}.

\emph{Kepler} is monitoring  nearly continuously the brightness of more than 150,000 stars selected from a 115 deg$^2$ field in the constellations  Cygnus, Lyra, and Draco with a 29.4 minute cadence.  An automated routine, the Transiting Planet Search (TPS) algorithm \citep{2012arXiv1201.1048T,2010ApJ...713L..87J, 2002ApJ...564..495J} searches the data for periodic signals which might indicate the presence of a planet. In order to avoid being swamped with false positives, TPS requires a phase-folded detection statistical confidence better than 7.1-$\sigma$, a threshold chosen such that the entire set of \emph{Kepler} observed stars could be searched within the first two years of the mission with less than one false positive transit detection owing to white noise \citep{2010ApJ...713L..87J, 2010ApJ...713L.109B}. Threshold-crossing events are further screened and those passing a series of data validation steps are assessed by members of the \emph{Kepler} science team, including individual visual inspection of each candidate light curve  \citep{2010SPIE.7740E..42W,2010ApJ...713L.103B}. Those that pass  become planet candidates promoted to \emph{Kepler} Object of Interest status (KOI). The \emph{Kepler} team's detection and  validation process have proved to be remarkably effective, producing more than 1235 planet candidates orbiting 977 stars \citep{2011ApJ...736...19B} (which would double the number of known planets), including 170 multiple planet (multiplanet) systems \citep{2011ApJS..197....8L}, the first detection of Earth-sized planets \citep{2011arXiv1112.4550F, 2011arXiv1112.4514G},  and the detection of unusual systems, such as the discovery of circumbinary planets \citep{2012Natur.481..475W,2011Sci...333.1602D}. 

Only a few percent of these candidate discoveries have been confirmed to date via radial velocity measurements, transit timing variations, or the statistical rejection of false positives that mimic transit signals  \citep[e.g.][]{2011arXiv1112.4550F, 2011arXiv1112.4514G,2011arXiv1112.1640B,2011arXiv1112.2165H,2011arXiv1110.5462B}. The majority of the \emph{Kepler} target stars are 12-16th magnitude, most  too faint to enable efficient radial velocity follow-up of likely candidates using even the largest ground-based telescopes. The  bulk of the KOIs will therefore likely remain unconfirmed. However, the \emph{Kepler} planet candidate sample nonetheless allows a statistical assessment of the quality of the remaining candidates to be made, and for limits on the likely planetary population of the Milky Way to be derived.  For example, a statistical analysis of \emph{Kepler} candidates with periods of less than 50 days around Solar-type stars using a 10-$\sigma$ signal detection threshold was performed by \cite{2011arXiv1103.2541H}. The frequency with which stars have planets is found to increase with orbital period and decrease with radius as expected from core accretion models. The effective temperature of the star was also found to be an important factor, with cool stars more likely to host planets. 

These results naturally depend on the performance of the \emph{Kepler} transit detection and vetting processes. There remains a pressing need for an independent assessment of the ability of the combined automated TPS  pipeline and data validation process to recover candidates. The \emph{Kepler} light curves are complex, with many exhibiting significant structure which includes multiple oscillations as well as quasi-periodic oscillations and short-lived variations. Traditional transit searches and validation techniques looking for a periodic signal may potentially miss transit signals that are dominated by the natural variability of the star or the \emph{Kepler} light curves may have  artifacts or other systematics  that might potentially compromise automatic fitting routines.  We employ a novel method of searching for transit signals in the \emph{Kepler} light curves via visual inspection. The human brain is a superb pattern recognition device and with minimal training can often outperform the most sophisticated machine learning devices and perform tasks not easily adapted to computer algorithms \citep{2012arXiv1201.6357S,21926992,2010arXiv1008.5387D,2009MNRAS.399..129L,2008MNRAS.389.1179L,2008ASPC..389..219M}. The detection of transits is essentially an exercise in pattern recognition, and is thus ideally suited to human classification. Transits that may be missed by traditional transit detection and vetting techniques may be identified by the human eye; highlighted by the serendipitous detection of a previously unknown transiting brown dwarf found by visual inspection of the Kepler  light curves \citep{2011ApJ...730...79J}. 

While classifying light curves from 150,000 stars would be beyond any single classifier, by using the World Wide Web it is possible to gather multiple independent assessments of the same data set. This approach also allows for a careful analysis of the reliability of any individual classification. With Planet Hunters\footnote{http://www.planethunters.org} \citep{2012MNRAS.419.2900F},  part of the Zooniverse\footnote{http://www.zooniverse.org} collection of citizen science projects \citep{2011MNRAS.410..166L,2008MNRAS.389.1179L}, the \emph{Kepler} light curves have been reviewed by eye for the signatures of extrasolar planets collectively by more than 173,000 volunteers to date with over 24,000  citizen scientists participating in the analysis presented here. Although novel in the search for exoplanets, human classification and crowd-sourcing citizen science have been applied to a wide range of astronomical applications  including the morphological classification of galaxies  \citep[e.g.][]{2011MNRAS.410..166L, 2008MNRAS.389.1179L,2008MNRAS.388.1686L}, the search for supernovae \citep{2011MNRAS.412.1309S}, the identification of solar coronal mass ejections \citep{2012MNRAS.420.1355D}, cataloging of star forming regions \citep{2012arXiv1201.6357S}, and cratering counts on the Moon \citep{2011A&G....52b..10J} in the Zooniverse suite of projects, as well as for discovery of solar grains entrapped in aerogel collected by the Stardust mission \citep{2008ASPC..389..219M}, and cratering counts on Mars \citep{2008ASPC..389..219M}.

The success of the Planet Hunters has already been demonstrated by \cite{2012MNRAS.419.2900F}, who report, in a preliminary search of \emph{Kepler} Quarter 1 light curves, the discovery and follow-up of two new planet candidates which were not identified as planet candidates in the \emph{Kepler} team's analysis of the first four months of \emph{Kepler} observations. Transits in the light curves of KIC 10905746 and KIC 6185331 were detected by TPS but failed to pass the data verification stage and therefore not promoted to KOI status \citep{2012MNRAS.419.2900F}. These discoveries, interesting in themselves, illustrate the desirability of an independent check on the \emph{Kepler} team's planet transit detection and validation scheme.  

This paper presents an independent assessment of the \emph{Kepler} planet inventory, examining the abundance of  $\ge$ 2 R$_{\oplus}$ planets on short period  ($<$ 15 days) orbits in the first 33.5 days of \emph{Kepler} science observations (quarter 1), based on the classifications gathered from Planet Hunters. We detail how the classifications were obtained and processed and present our candidate detection scheme. We also assess the detection efficiency of the Planet Hunters project including a comparison to the \emph{Kepler} inventory of planet candidates. 

\section{\emph{Kepler} Quarter 1 Data}

The \emph{Kepler} Quarter 1 (Q1) dataset consists of observations for 156,097 stars spanning 2009 May 13$-$2009 June 15. The data were obtained from the \emph{Kepler} Mission Archive\footnote{http://archive.stsci.edu/kepler} hosted by the  Multi-Mission archive at STScI (MAST\footnote{http://archive.stsci.edu}). The majority of the light curves used in our analysis are from the 2010 June 15 data release. A subset of Q1 targets, 5828 stars, had long cadence light curves publicly released after 2010 June. We obtained data for those stars from the 2011 February 2 release. Due to an error in uploading, five stars (KIC 8349808, 3656476, 6603624, 11446443 [KOI 1.01  \citep[Tres-2,][]{2006ApJ...651L..61O}],  KIC 11026764), comprising $\sim$0.004$\%$ of the total \emph{Kepler} sample, were not included in this analysis. 

Planet Hunters utilizes the long cadence observations from \emph{Kepler} where measurements are effectively obtained once per 29.4-minute interval, presenting the scaled systematic error-corrected flux outputted by the \emph{Kepler} science processing pipeline \citep{2010ApJ...713L..87J}. The flux values and uncertainties used correspond to the Pre-Search Data Conditioning (PDC) PDCSAP\underline{\hspace*{0.25cm}}FLUX  and PDCSAP\underline{\hspace*{0.25cm}}FLUX \underline{\hspace*{0.25cm}}ERR, previously labeled AP\underline{\hspace*{0.25cm}}CORR\underline{\hspace*{0.25cm}}FLUX and AP\underline{\hspace*{0.25cm}}COR\underline{\hspace*{0.25cm}}ERR columns of the released \emph{Kepler} FITS tables \citep{Keplerarchivemanual}. The PDC correction removes systematic effects and long term trends  from the light curves. For further details of the \emph{Kepler} Q1 observations and data processing we refer the reader to \cite{2011ApJ...728..117B} and \cite{2010ApJ...713L..87J,2010ApJ...713L.120J}.

\section{Planet Hunters}
\label{ref:PH}
\begin{figure}
\epsscale{1}
\plotone{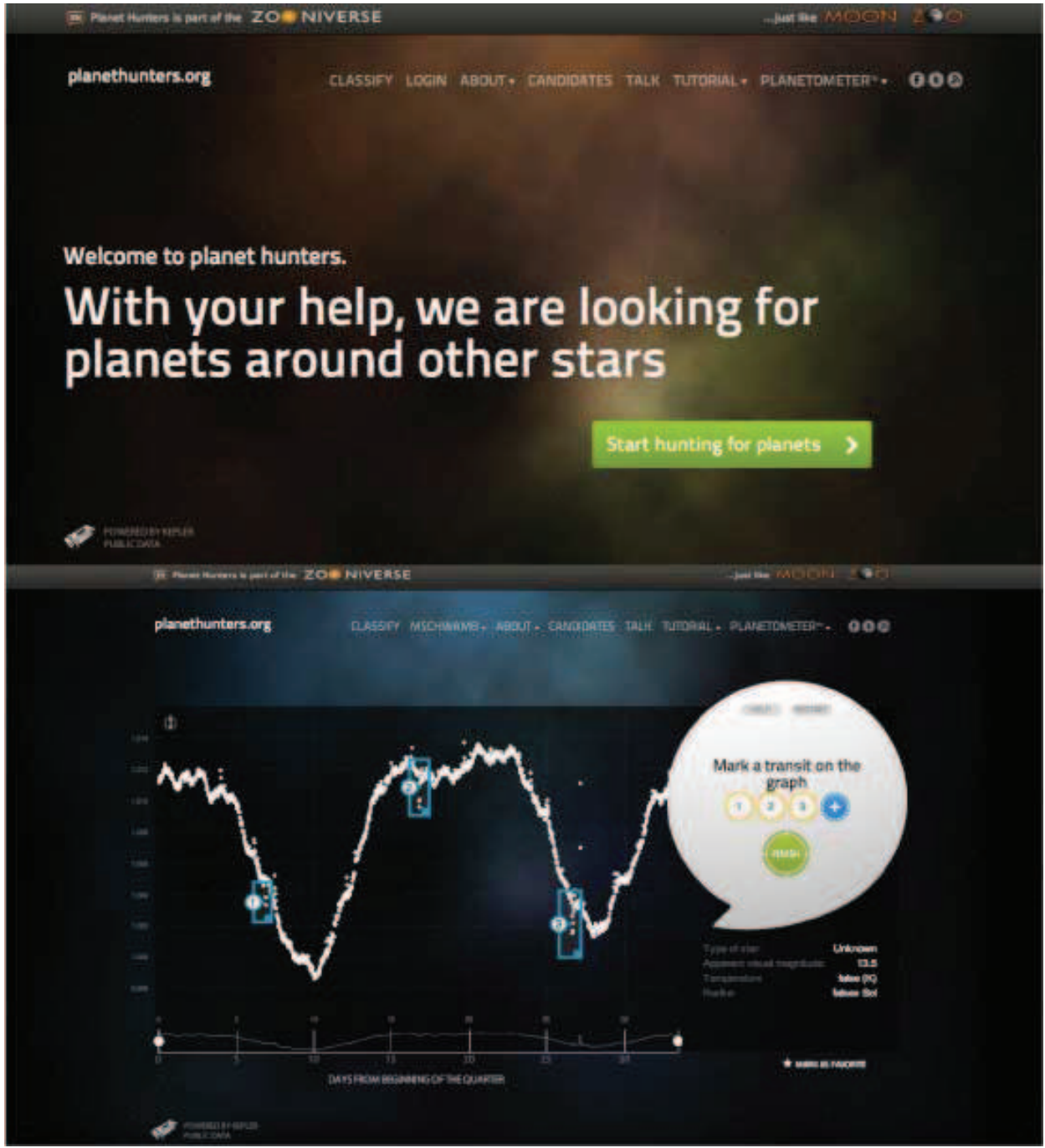}
\caption{ Front page (top) and main classification interface (below) of the Planet Hunters website. }
\label{fig:fig1}
\end{figure}

The Planet Hunters interface presents the volunteer with a single 33.5-day Q1 light curve for a given star, plotting the star's relative brightness (Figure \ref{fig:fig1}). Error bars representing the 1-sigma measurement uncertainties are also displayed in the interface. The default setting is to show the entire 33.5 days at once. Volunteers have the ability to freely zoom the time axis or to scan the curve while zoomed in. A button allows the zoom on the y-axis (relative brightness) to automatically be adjusted to an appropriate level. The y-axis range spans from the minimum value to the mean level plus three times the standard deviation. The display is then adjusted to 5-85$\%$ of that range; this was found experimentally to produce a good result for the majority of curves.

Visitors to the Planet Hunters site can begin classifying almost immediately. Basic training is provided via an inline interface, accessible with a single click from the Planet Hunters home page. A sample light curve with example simulated transits is shown before volunteers can review real light curves. More detailed training material is also available elsewhere on the site. It is possible to train and then to take part in the project without registering a Zooniverse account, although volunteers who are not logged in will receive repeated reminders to register. The site is designed so that volunteers are unable to select which curve to classify in order to minimize the risk of deliberate manipulation of input data.  Light curves are prioritized according to the stellar properties and previous classifications as described in \cite{2012MNRAS.419.2900F}. To ensure the volunteers have no prior knowledge about a given light curve that may effect their judgment, the KIC identifier is not revealed during classification; only an approximation of the star's stellar temperature, radius, and magnitude are presented to the reviewer. 

\begin{figure}
\epsscale{0.7}
\plotone{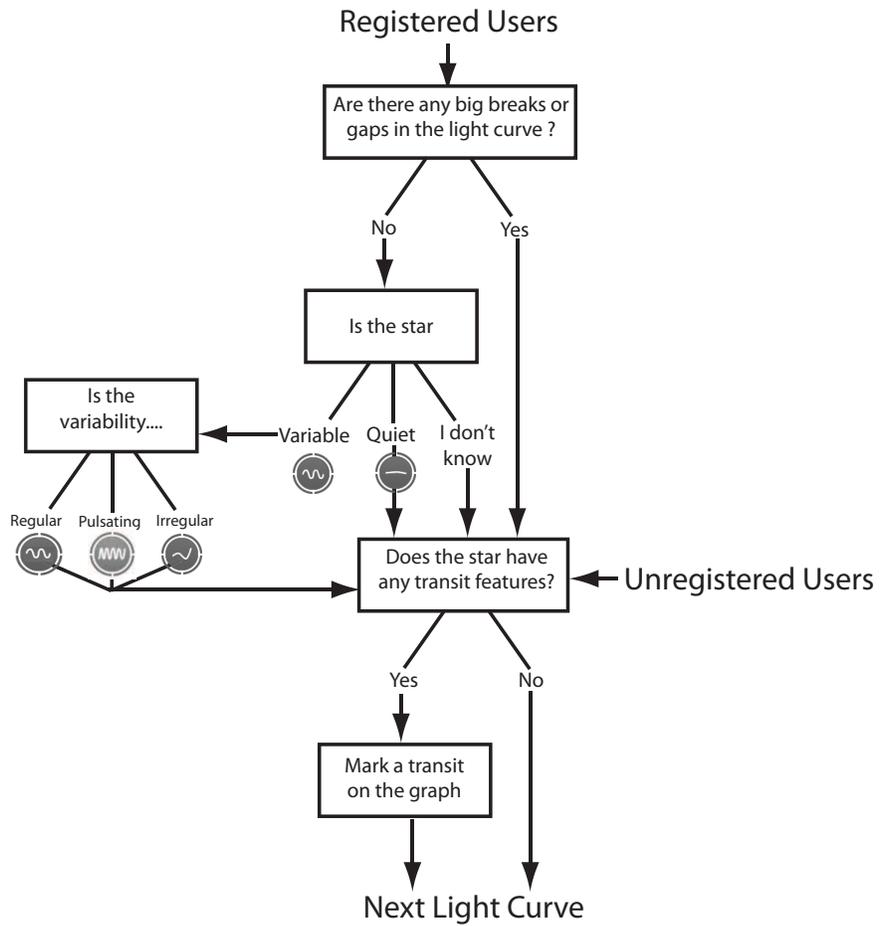}
\caption{ The decision tree that a Planet Hunters classifier is presented with when reviewing a light curve.  }
\label{fig:fig2}
\end{figure}

\begin{figure}
\epsscale{1}
\plotone{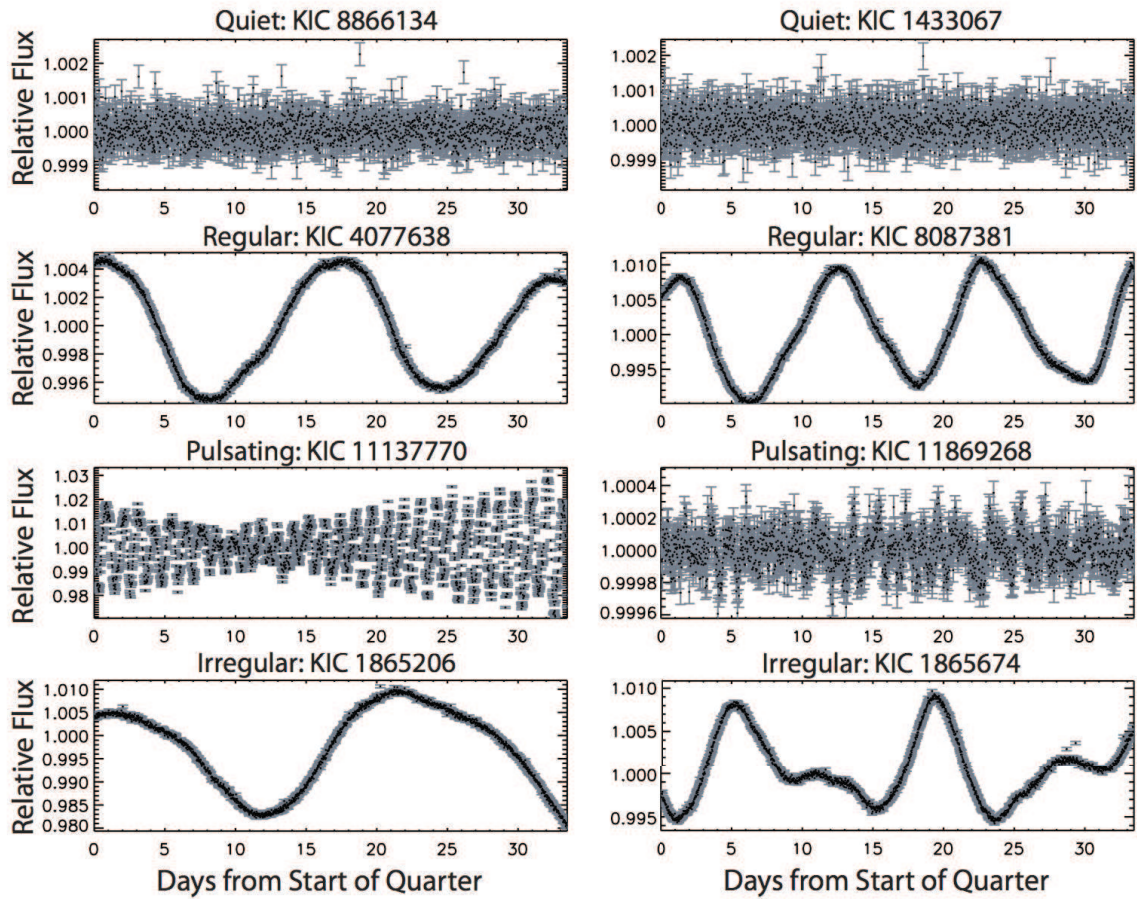}
\caption{ Examples light curves in each class. Reported error bars are overplotted in gray. For each light curve presented 75$\%$ or greater of the classifications selected the specified type.  }
\label{fig:fig3}
\end{figure}

Classification proceeds via a decision tree, as detailed in Figure \ref{fig:fig2}. The first series of questions produce a coarse categorization of the light curve's variability which is primarily intended to assist in identifying transits and in identifying any gaps or defects  in the light curve that may have been introduced during the PDC process, typically due to removing a cosmic ray event (Jenkins et al 2009).  If the light curve has no visible gaps, volunteers are asked to categorize the star as `variable' or `quiet'.  If the star is deemed variable, the volunteer is further prompted to characterize the variability as: `regular' (showing periodic repeating pattern), `pulsating' (rapid variations observed in star's brightness on the timescales of a few hours to days), or `irregular' when no discernible pattern is seen. Representative light curves for each variability class are presented in Figure \ref{fig:fig3}. 

Volunteers are then asked to mark any transit features visible in the light curve. Clicking a button produces a box on the light curve (Figure 1) positioned  in the center of the review interface  which can then be moved and resized. If necessary, more than one box can be added to each curve, and transits marked in error can be removed.  The current zoom level of the interface determines the default box size. The box width and height initially correspond respectively to 1/8th the x-axis range and half the y-axis range displayed. A second version of the interface, introduced on 2011 Jan 31, allows in addition for transit boxes of any size to be drawn directly onto the curve, without the need to click a button first.  In the upgraded interface, if a transit box already exists when the button to add a new transit box is clicked, a box with the same dimensions as the one previously made is generated. To aid in refining the transit box dimensions, with a single click the improved interface will automatically center on a selected transit box, zooming in to a width twice the size of the region highlighted by the specific transit box. Due to the way that transit boxes are presented on the screen, only approximately eighteen boxes can be placed without hiding the button to submit the classification, effectively creating a limit on the number of transits that can be identified. This issue only significantly affects the marking of transits from planets with orbital periods less than $\sim$2 days. 

Classifications are stored in a live Structured Query Language (SQL) database. For each entry question in the decision tree, the time stamp, user identification, light curve identifier, and response are recorded. If transit boxes are drawn, the position of the box center, width, and height of each box are recorded as well.  Each of the 156,092 Q1 light curves is examined and classified multiple times, by normally 10 and at least 5 classifiers including both logged-in and unregistered users before being retired and removed from circulation. The variability and gap questions are only asked of those volunteers registered with a Zooniverse account; non-logged in classifiers are asked only to identify transit-like features.  Before the first classification and subsequently after every 5 classifications, a non-logged classifier is reminded to create a Zooniverse account. If the user registers, subsequent classifications are attributed to the user in the classification database, but previously made classifications are labeled as from a non-logged in user.  

Once the classification is submitted, the Planet Hunters internal designation for the light curve is revealed to the user. Classifiers are then asked if they would like to discuss the particular light curve on the Planet Hunters Talk site\footnote{http://talk.planethunters.org - The code is available under an open-source license at https://github.com/zooniverse/Talk}. Each Q1 light curve has a dedicated page where users can write comments, add searchable Twitter-like hash tags, and link similar light curves together. This system is useful in identifying objects of particular interest and unusual light curves that are difficult to find via automatic classification schemes, a complement to the systematic analysis described here.

110,997(71$\%$) of the \emph{Kepler} Q1 light curves  were completed with the original version of the interface and decision tree launched on 2010 December 16. We note that on 2011 Feb 1, the decision tree was modified for the inclusion of \emph{Kepler} Quarter 2 data. All Quarter 2 light curves  presented in the Planet Hunters interface have missing data gaps  due to safe mode events or the \emph{Kepler} spacecraft pointing towards Earth to transmit data. In the original interface, the light curve variability questions were skipped if the user had identified a gap in the light curve. The decision tree was modified  such that if there was a gap in the light curve the variability questions would still be asked. In subsequent changes on 2011 March 9, the gap question was removed entirely. The responses to the gap question are not used in identifying possible planet candidates.  7,759 light curves received classifications after 2011 May 1 when the majority of the light curves presented in the classification interface were Quarter 2 \emph{Kepler} light curves. The appearance and quality of the Quarter 2 observations is different from that of Quarter 1 which may make Q1 light curves identifiable to more experienced users. We address the introduction of Q2 light curves on Q1 classifications  and the impact of the interface modifications on the user responses and behavior in Appendix \ref{ref:biases}. 

\subsection{Synthetic Transits}

In order to characterize our detection efficiency, simulated transit events were injected into a subset of the \emph{Kepler} light curves and presented on the Planet Hunters classification interface. Synthetic transit light curves were created from the Q1 light curves randomly sampling from the 156,097 publicly released long cadence light curves excluding known \emph{Kepler} planet candidates \citep{2011ApJ...736...19B}, transit false positives \citep{2011ApJ...736...19B}, and known eclipsing binaries  \citep{2011AJ....142..160S,2011AJ....141...83P} available online as of 2011 April\footnote{http://keplerebs.villanova.edu/}.  A total of 6494 synthetic light curves were created. Simulations were divided into 7 radii bins from  2-15 R$_{\oplus}$ and  6 orbital period bins between 0.5 and 15 days. Table \ref{tab:sims} details the number of synthetic light curves generated for each radii and period bin.  

Synthetic transit points are implanted into the \emph{Kepler} light curve by subtracting the amount of flux decremented during the transit, following the equations of  \cite{2002ApJ...580L.171M} and \cite{2010arXiv1001.2010W} for transit depth and duration and assuming no limb darkening. We neglect inclination and eccentricity effects assuming circular edge on orbits  as viewed from Earth. We do not adjust the measurement flux errors for data points implanted with the transit event, providing a conservative and slight overestimate of the Poisson measurement noise. The starting time of the first injected transit is randomly chosen between the starting time of the light curve and one orbital period later. Each simulation has at least two injected transits present in the light curve.

We adopt the stellar radii and surface gravity (i.e. log(g)) values reported in the \emph{Kepler} Input Catalog\footnote{http://archive.stsci.edu/Kepler/Kepler\underline{\hspace*{0.25cm}}fov/search.php} (KIC, \citeauthor{2011AJ....142..112B} \citeyear{2011AJ....142..112B}) to compute the stellar mass and the transit depth. The KIC uses multi-band color photometry to distinguish between main sequence dwarf  and cool giant stars in the \emph{Kepler} field. The KIC performs this task reasonably well  with only a small fraction of the catalog  estimated to be giants misidentified as dwarf stars \citep{2011AJ....142..112B, 2010ApJ...713L..79K}. The KIC provides a first estimate of the stellar properties of the \emph{Kepler} target stars, with log(g) reliable to within 0.4 dex with a 0.25 dex rms in comparison to high resolution spectra taken for a subsample of KIC stars \citep{2011AJ....142..112B}. The reported uncertainty for the KIC stellar radii varies. \cite{2011AJ....142..112B} note that subgiants misidentified as main sequences stars may have their radii underestimated by as much as a factor of 1.5-2. According to \cite{2011arXiv1103.2541H}, stellar radii are reliable within an rms of 35$\%$, with \cite{2011ApJ...736...19B} quoting errors in stellar diameters as high as 25$\%$.  \cite{2011ApJ...738L..28V} find that the KIC systematically underestimates stellar radii by as much as 50$\%$ for stars with radii less than 1 R$_{\odot}$. Given this large uncertainty in the  KIC values and differences in the reported error estimates, we assume the KIC is representative of the bulk properties of the \emph{Kepler} stars and simply employ the KIC calculated stellar parameters in the same manner as other previous works   \citep[e.g.][]{2011arXiv1103.2541H, 2011ApJ...736...19B, 2011ApJ...732L..24L}.

The simulations are presented to logged-in users only and classified in the exact same manner as the real light curves. After the classification is submitted, the user is notified that the light curve was simulated with the injected transit points identified. We note that the simulations were classified on the Planet Hunters site after the completion of the majority of the Q1 light curves and were presented alongside the Quarter 2 light curves in the second version of the classification interface. Quarter 2 is substantially and notably different in data quality than Q1 \citep{Keplerdatamanual}. However, we expect these effects to be small and the complete simulation set to be a good test of volunteer behavior on the entire Q1 data set. We discuss the impact of these effects and detection efficiency further in Appendix \ref{ref:biases}.

\section{Selecting Transit Candidates}

In this section we detail our algorithm for combining the individual classifications received for each light curve and the selection pipeline developed to produce a final catalog of transit candidates. Each light curve in our sample was examined and classified multiple times, with a minimum of 5 and a mean of 10 independent  classifications. In total, 1,687,547 individual classifications were generated searching the Q1 data and examining the synthetic light curves. 24,300 registered volunteers made 98$\%$ of those responses, with the remaining 2$\%$ from unregistered volunteers. Figure \ref{fig:fig4} plots the distribution of user classifications for registered users. As is common for online projects \citep{Zachte2012,Crowston2008c}, most volunteers typically classify only a few light curves, a median of 5 and mean of 68 light curves. 47$\%$ of logged-in users have classified fewer than 5 Q1 light curves. Only 10$\%$ of classifiers have made more than 100 classifications. 

\subsection{User Weighting}

Rather than combining results from all classifiers equally in a majority vote, the first step in converting the raw classifications to transit candidates is to determine a user weighting.  The weighting scheme distinguishes those classifiers who are more sensitive to identifying transits and less prone to providing false positives in order to consider their responses more heavily than others when selecting transit candidates. We evaluate the ability of each classifier and assign a weight based on their tendency to agree with the majority opinion and for those who classified the synthetic light curves, on their performance identifying the simulated transit events. We follow a modified prescription developed by \cite{2008MNRAS.389.1179L} and \cite{2008MNRAS.388.1686L} for user weighting based on the visual morphological classifications of galaxies from the Galaxy Zoo project.

We assign user weights in a two stage process. Initially, all users start out equal, with a weight of unity. We then use the results of the synthetic light curves to seed the initial weighting for those users that classified a synthetic light curve. From the synthetic transits we have a direct evaluation of an individual user's ability to identify transits for a given planet radii and orbital period. Only logged-in users are served synthetic light curves to classify. 6,307 (26$\%$ percent) of the registered Q1 volunteers examined at least one synthetic light curve.  A median of 2 and a mean of 10 simulations were classified by those 6307 volunteers. For every synthetic light curve and classifier who viewed the simulation, we evaluate how well the user identified the injected transits events. The synthetic transit is deemed ÒmarkedÓ if the middle of the injected event is within one of the user boxes drawn. A given user-drawn box only counts once.  If the user draws a single box around two simulated transits, credit is only given for successfully identifying one transit. 

For each simulation a user classifies, his or her weight is incremented by the fraction of markable transits  correctly identified. Typically the number of markable transits is the number of synthetic transits present in the light curve but  for those few simulations with more than 18 transits (only 14$\%$ of the total) where marking more would cause the submit button to disappear effectively limiting the transits that can be identified, the maximum number of markable transits is set to 18, and the weight of any volunteer who correctly marked 18 or more transits is increased by one. If no synthetic transits are marked, the user weight is decreased by 0.2 times the fraction of the simulation's classifiers who marked at least one transit on the simulated light curve.  User weights are normalized such that the maximum user weight attainable, if all synthetic transits seen were identified correctly, is 2.  Thus the weighting scheme rewards more heavily users who mark correctly the majority of the injected transits and  down-weights  users more significantly for missing easily identifiable transits and less so for harder to perceive transit depths.

At this point, user weights have only been modified for those logged-in users who have classified at least one synthetic light curve. All other registered volunteers still have an initial user weight of 1. We next use an iterative scheme to assign weightings to all users based on the responses for all the real Q1 light curves. We divide user responses into two classes: `transit' or `no transit'  based only on whether or not a user drew a transit box. For a given \emph{Kepler} star \emph{i} and for each class \emph{j}, a score $s_i(j)$, is then calculated. We define $s_i(j)$ as:
\begin{equation}
s_i(j)=\frac{1}{W_i } \displaystyle \sum_{k} w_k \qquad    k \textrm{=users who voted \emph{j} for light curve \emph{i}}
\end{equation}
where $W_i$ is the sum of the users weights for all the users who classified light curve \emph{i}. We then update $w_k$ , the weight of user \emph{k} based on the calculated light curve transit scores ($s_i(j)$). A user whose behavior is in line with the majority  weighted vote will be rewarded and up-weighted, whereas those that disagree with the majority opinion will be down-weighted.  We adjust $w_k$ as follows:
\begin{equation}
w_k = \frac{A}{N_k}  \displaystyle \sum_{i} s_i (j \textrm{ for light curve } i \textrm{ chosen by user } k) \qquad    i \textrm{= all light curves reviewed by user \emph{k}}
\end{equation}
where $N_k$ is the number of Q1 light curves classified by user \emph{k}, and scaling factor $A$ is chosen to be such that the mean user weight  remains at 1. 

The light curve scoring and updating of the user weights is iterated until convergence is achieved, when the median absolute difference between the old and updated user weights is less than or equal to 1x10$^{-4}$. Classifications from non-logged in users are included in the weighting scheme to calibrate the other logged-in users but all non-logged in users are effectively treated in the weighting scheme as a single user with an unchanging weight of 1. 
All users have final weights greater than zero with 55$\%$ of the logged-in users having values greater than or equal to 1. The final distribution of user weights is plotted in Figure \ref{fig:fig5}. 

\subsection{Candidate Selection}

Once a user weighting solution has been found, the Q1 light curves are searched for transit candidates. In this analysis, our search focuses on the identification of planet candidates with radii greater than 2  R$_{\oplus}$ and orbital periods less than 15 days. From this point forward, the synthetic light curves are treated in exactly the same as the real \emph{Kepler} light curves. Table \ref{tab:remaining} summarizes the candidate selection process and the number of light curves and simulations remaining at each step.  

\subsubsection{Transit Score} 
We use the `transit' scores, $s_i(\textrm{transit})$,  for each of the \emph{Kepler} light curves as a measure of the light curve's likelihood to have a planet transit.  Each synthetic light curve is scored using the previously calculated user weights to determine $s_i(\textrm{transit})$. We plot the cumulative distribution of transit scores for the Q1 \emph{Kepler} light curves and simulations in Figure \ref{fig:fig6}.  

We aim to include as many possible transit candidates without the number of false positives light curves overwhelming the list. A cut in $s_i(\textrm{ transit})$ is made to create the initial list of potential transit candidates. This threshold was chosen after visual inspection of a random subset of the remaining light curves and light curves with transit-like events detailed in  \cite{2012MNRAS.419.2900F}  previously found during a preliminary search through the Q1 data set.  The science team found that at $s_i(\textrm{ transit})$ $=$ 0.5 the false positives were overwhelming the number of candidate transit detections. Therefore we select our potential planet candidates as those light curves with $s_i(\textrm{ transit})$ $>$ 0.5. 5,511 (3.5$\%$) of the \emph{Kepler} light curves have $s_i(\textrm{ transit})$ greater than 0.5. 

\subsubsection{Removing Single Transit Events}

We restrict our search to transiting planets with periods less than 15 days where there will be at least two transit events present in Q1.  The 50$\%$ transit score threshold only discriminates between the likelihood of the light curve having a transit or not, and is not dependent on the number of transits present in the light curve. There will inevitability be light curves with single transit-like events due to longer period transiting planets (orbital periods greater than 15 days) and primary transits from eclipsing binaries with orbital periods greater than 15 days where the secondary eclipse will occur in future quarters of \emph{Kepler} data. We remove the most obvious of these single transit-like events from our candidates list, discarding light curves where a single transit box was  drawn by all classifiers who identified transits. Of the 5,508 candidates, 570 light curves were removed by this cut. These cut light curves were separately analyzed for potential transits  and the first candidates from that analysis are further discussed in \cite{2012arXiv1202.6007L}.

\subsubsection{Light Curve Variability}

Visual inspection of the remaining candidate light curves revealed that a significant portion were pulsators, light curves which exhibit rapid variations in the star's brightness on the timescales of a few hours to days. Many classifiers mistook the individual pulsations for transits. To further cull the candidates list of these false positives, we assess the light curve variability on timescales less than 33 days, placing a more stringent transit score threshold for the pulsating light curves.

While classifying, users are asked to broadly characterize the observed variability of the Q1 systematic error-corrected light curves as described in Section \ref{ref:PH}. We note this may not reflect a true assessment of the stars' intrinsic astrophysical variability. The Pre-Search Data Conditioning (PDC) applied to the raw \emph{Kepler} light curves may remove real astrophysical signals from, and potentially inject high-frequency signals, into the light curves when correcting for instrumental systematics and optimizing the light curves for transit detection \citep{2010vC, 2010ApJ...713L..87J}. The Planet Hunters variability assessment serves as an evaluation of the bulk properties of the Q1 light curve sample rather than the properties of individual \emph{Kepler} target stars.

We first group the light curves into ÒquietÓ and ÒvariableÓ classes, then further subdividing the variable stars into ÒpulsatingÓ, ÒregularÓ, and ÒirregularÓ. We combine together the multiple user classifications by simply taking the majority, selecting the variability class with the highest fraction of votes.  If the light curve has less than 3 classifications or a majority is not reached, we deem the light curve unclassified. Table \ref{tab:variability} summarizes the variability properties of the Q1 light curves based on Planet Hunters classifications. The majority of the \emph{Kepler} light curves, 65.6$\%$ are quiet with the bulk of the variable stars being classified as pulsators (48.9$\%$ of variable light curves, 16.5$\%$ of the full sample). 

The majority answer does not necessarily mean the proper identification of the light curve's variability. Little direct instruction is provided to the user when categorizing the light curve's variability. The inline tutorial does not address the variability questions asked in decision tree, instead focusing on the identification of planet transits, but there are prototypes for each variability class in the help material that is accessible to the user by a single click during classifying.   \cite{2011arXiv1111.5580M}, \cite{2011AJ....141..108C}, and  \cite{2010ApJ...713L.155B}  assess the photometric variability of the Q1 dataset to which we can compare. \cite{2011arXiv1111.5580M}  and \cite{2010ApJ...713L.155B}   examine the raw pre-PDC processed light curves applying their own methods to remove systematic effects while preserving stellar variability signals, and \cite{2011AJ....141..108C} examine the of variability the post-PDC processed light curves. Because users classify the PDC light curves,  we choose to evaluate the reliability of the Planet Hunters assessments for the \emph{Kepler} Q1 dataset with \cite{2011AJ....141..108C}, where we can make a direct comparison.

\cite{2011AJ....141..108C} divide the variability from a sample of 126,092 dwarfs and 17,129 giants using a reduced chi-squared threshold of 2. Although  \cite{2011AJ....141..108C} group the light curves by spectral type and into dwarfs or giants, we report the total fractions for the entire sample. In Table \ref{tab:varcompare} we compare the variability fractions from the \cite{2011AJ....141..108C} to this work.  Like \cite{2011AJ....141..108C} we only examine stars that have KIC entries for radius and temperature. We observe the same trend of light curve variability decreasing with increasing magnitude with 49.6$\%$ of stars with instrumental magnitudes less than 14 in our sample deemed variable. This is to be expected as measurement precision is highest for the brightest stars in the sample. 

The variability fractions of the light curves with transit scores greater than 0.5 are detailed in Table \ref{tab:varcand}. Only 497, or 10$\%$,  of the candidate light curves had too few classifications to secure variability identifications. The pulsators are overrepresented in the candidates, composing 36$\%$ the sample, nearly double the amount in the entire sample as a whole, confirming the results from visual inspection of the candidates. It is not surprising that the largest source of false positives in our candidates list would be due to pulsators.  With the user classifications producing an accurate picture of Q1 light curve variability, we can use the Planet Hunters variability identifications to place a more stringent threshold for pulsators and remove pulsator light curves with transit scores less than 0.8 from the remaining list of planet candidates.  Of the 4938 candidates, 1534 light curves  were removed by this variability cut.  

\subsubsection{Round 2 Review}

After the series of cuts described above, 3404 candidate light curves and 4974 simulations remain, including a mixture of eclipsing binaries \citep{2011AJ....142..160S, 2011AJ....141...83P}, \emph{Kepler} planet candidates \citep{2011ApJ...736...19B}, transit false positives, and potentially previously unidentified planet candidates. Planets with periods less than 15 days will have at least two transits during the Q1 observations. In the analysis presented here, we are only interested in those light curves with two or more identified transits. The current selection criteria only determine whether or not transits are likely to present in a given light curve, not identifying the number of visible transits found by the Planet Hunters classifiers. In order to sort through the remaining list of candidates and identify those light curves with at least two transits, a second classification interface was built and second round of review performed. 

In this simplified interface, the full 33.5-day light curve is plotted with the same horizontal and vertical range as presented in the main classification interface To assist the reviewer, the positions of all the user-drawn transit boxes are overlaid on the light curve in blue such that regions where there is more overlap between the transit boxes are shaded darker.  Reviewers can toggle on and off the plotting of the transit box locations with a click of a button. The light curve is presented as a static image; Round 2 reviewers are not able to zoom in or adjust the resolution of the light curve like the main classification interface. Both real and simulated light curves are presented  and reviewed blindly by the screeners. Unlike the main classification interface, simulated light curves are not identified after classification. 

Screeners for the Round 2 review were recruited from the  Planet Hunters Talk site and through a blog post. These volunteers are asked if the presented light curve has two or more transits. The volunteer has three possible responses: `yes', `no', or `maybe'.  In Round 2, the reviewers are not asked to determine whether the transits are due to a single planet or to identify possible secondary eclipses within the light curve. Instead we instruct the user to ignore whether the transits have the same depth in the tutorial information and ask them to focus only on the number of transit-like events identifiable in the light curve.  

Round 2 review was performed in two stages; the first in September 2011 and the second in February 2012. A much smaller number of classifiers examined the light curves in Round 2 review interface compared to the total number of individuals who reviewed Q1  light curves and simulations in the main Planet Hunters interface. 137 volunteers participated in Round 2 review. Every candidate light curve and simulation received at least 5 and an average of 9 independent classifications. A total of 75,291 classifications were made in the Round 2 review interface with volunteers examining an average of 550 and median of 75 light curves and simulations. The majority of Round 2 classifiers are experienced users, having classified a median number of 177 Q1 classifications, with 55$\%$ of the Round 2 reviewers having classified more than 100 Q1 light curves and simulations.  Only 25  of the Round 2 reviewers had not previously classified a Q1 light curve. 72$\%$ of the classifiers have user weights greater than or equal to 1; therefore, we choose to simply tally the number of responses for each answer choice (`yes', `no', `maybe').  If the candidate light curve received a majority of `yes' votes, the light curve is passed on for further review and scrutiny by the Planet Hunters science team.  

\subsection{Planet Candidates}

In total 1220 light curves, 0.78$\%$ of the Q1 light curves, were passed on for further inspection. We note one light curve, KIC 8953426,  which is a known eclipsing binary  \citep{2011AJ....142..160S, 2011AJ....141...83P},   did not receive Round 2 review due to a processing error but was passed on with the other remaining candidates to inspection by the Planet Hunters science team. At this stage, we identify and remove reported eclipsing binaries \citep{2011AJ....142..160S, 2011AJ....141...83P}, sources with an eclipsing binary contaminating the aperture listed in \cite{2011AJ....142..160S}, and additional planet transit false positives identified by  \cite {2011ApJ...736...19B} and \cite{2012MNRAS.419.2900F}. Those remaining which are not listed as KOIs as of February 2011 \citep{2011ApJ...736...19B} are visually inspected.

The Planet Hunters science team reviewed the 75 remaining  light curves, first verifying there were at least two transits present in the light curve. For those candidates with multiple transits visible, the Quarters 2-6 light curves were next inspected to confirm repeat transits in subsequent quarters. We note that Quarters 2-5 light curves are currently being analyzed by the Planet Hunters community. Odd/even transits were checked for depth to remove obvious unlisted eclipsing binaries. We also examined the Kepler target stars surrounding each of the remaining light curves for nearby  eclipsing binaries that could be contaminating the photometric aperture and mimicking transit-like signals in the candidate light curve. If the  transit times are  correlated with that of a known eclipsing binary light curve in the vicinity or after examination of the individual pixel by pixel light curves for the star with PyKE\footnote{http://keplergo.arc.nasa.gov/PyKE.shtml} it can be seen that only a small subset of pixels exhibit the transit features, the candidate is deemed a false positive. Table \ref{tab:round2} details the outcome of the review. 23 sources, including a known RR Lyrae star  \citep[KIC 6186029,][]{2010MNRAS.409.1585B}, did not have two or more visible multiple transit-like events in the Q1 light curve or were data processing glitches. 18  light curves did not have repeats in subsequent quarters of data, and 27 light curves have alternating transit depths indicating they are likely to be  previously unknown eclipsing binaries or have transits correlated with an nearby eclipsing binary contaminating the photometric aperture. This left only 7 surviving candidates (KIC 551108, 6616218, 8240797, 9729691, 10905746 11017901, 11551692) from our review not identified as KOIs in the \emph{Kepler} analysis of Quarters 0-2 \citep{2011ApJ...736...19B}. All 7 candidates have been subsequently identified in later runs of an updated version of TPS and data validation pipeline \citep{2012arXiv1201.1048T} and promoted to KOI status by the \emph{Kepler} team during a multi-quarter TPS analysis of Quarters 1-6 and data validation using products from Quarters 1-8 (Batalha 2012- private communication,Batalha et al. 2012).Additionally KIC  5511081, 6616218, 8240797, 9729691, 11551692  have been subsequently identified by the \emph{Kepler} team as multiplanet systems with at least one $\ge$ 1.9$R_{\oplus}$  planet candidate orbiting the host star in less than 15 days.

 All seven light curves are plotted in Figures \ref{fig:figcandidate_curves} and  \ref{fig:figcandidate_curves2}. 
 In Figure  \ref{fig:zoom} we provide a zoom-in of a chosen transit for each set of transits identified in the 7 light curves. Visually the science team could identify two separate sets of repeating transits in the mutli-planet KIC  8240797, 9729691, and 11551692 based on the user drawn boxes. We note that KIC 10905746 is a short period planet candidate first reported by \cite{2012MNRAS.419.2900F}, found in a preliminary search of the Planet Hunters Q1 classifications. The second planet candidate reported by \cite{2012MNRAS.419.2900F} was not in this analysis since it has a period of 40 days with only a single transit in Q1. Additionally  KIC 5511081 was also identified as a planet candidate by  Planet Hunters in an preliminary search of Q2 data as reported by \cite{2012arXiv1202.6007L} and \cite{2012arXiv1202.5852B}.
   
 We do not specifically know the reasons  why each of the 7 light curves failed to be promoted to KOI status in \citep{2011ApJ...736...19B}.   All 7 light curves were flagged as having potential transits with the multi-quarter TPS analysis of  Quarters 1-3  \citep{2012arXiv1201.1048T}, but this version of TPS was used to create the \cite{2012arXiv1202.5852B} candidate list and was not implemented during the  \citep{2011ApJ...736...19B} candidate selection. The \cite{2011ApJ...736...19B} version of TPS identified transits in individual quarters.  \cite{2012MNRAS.419.2900F} report that KIC 10905746  was identified by TPS but the automatic light curve fitting failed to converge to a  solution and the star was rejected. 4 of the 7 systems  have two transiting planets with periods less than 15 days, this may have been an additional factor. Regardless, for the exact reason these candidates were not initially identified, they are now KOIs by the multi-quarter TPS and upgraded data validation scheme detailed in  \cite{2012arXiv1202.5852B} and \cite{2012arXiv1201.1048T}. The lack of any other additional  candidates not currently KOIs highlights the effectiveness of the \emph{Kepler} detection scheme and the increased detection efficiency from improvements to the \emph{Kepler} detection pipelines.

\section{Detection Efficiency}

In this section, we discuss the detection efficiency of Planet Hunters for finding short period planets with radii larger than or equal to 2  R$_{\oplus}$ and periods less than 15 days.  We first assess the performance based on the detection of our sample of simulated transit light curves and then compare to the known sample of \emph{Kepler} planet candidates.  We explore the potential biases that may effect user responses and behavior and their impact on the calculated detection efficiency in Appendix \ref{ref:biases}. 

\subsection{Synthetic efficiency}

Simulations were processed and screened by Planet Hunters users in a similar manner to the real Kepler light curves, and Table \ref{tab:remaining} summarize the simulation light curves that were retained at each stage in the planet candidate selection process. Those synthetic light curves that survived the transit score threshold, the single transit removal step, and the variability cut  were shown in the Round 2 review. The simulations were shown with real light curves and at this stage users were not notified which light curves had injected transits or which ones were real \emph{Kepler} light curves.  

To estimate the detection efficiency, we bin the  simulations into the same radii and period bins used to generate the light curves. Figure \ref{fig:sims} (with Poissonian 68$\%$ errors as prescribed by \citeauthor{1991ApJ...374..344K}  \citeyear{1991ApJ...374..344K}) and Table \ref{tab:simsrecovery} present the recovery rates for the 6494 generated synthetic light curves. We deem a simulation recovered if the synthetic light curve successfully passed Round 2 review stage. We find that greater than or equal to 4 R$_{\oplus}$, independent of orbital period, Planet Hunters is $\ge$ 85$\%$ efficient at detecting short period planet transits. The detection efficiency begins to drop at 3-4 R$_{\oplus}$ to 75$\%$ and for the 2-3 R$_{\oplus}$ bin 40$\%$ of the synthetic light curves were recovered.  For the larger planet radii bins in Figure \ref{fig:sims}, the majority of the missed simulations were around $\ge$ 2 R$_{\odot}$ stars where the transit depths are as low as those for $<$ 3 R$_{\oplus}$ planets around solar radii stars. For all the radii bins, the detection efficiency was relatively insensitive to orbital period. We note that we injected into a random sampling from the Q1 light curves including giant stars, thus our synthetic recovery rate represents the Planet Hunters detection efficiency for the overall \emph{Kepler} target star sample. 

We find that our detection efficiency is independent of the number of transits present in the light curve. Classifiers are are just as likely to identify and mark transits in  a light curve if there are 2 or 18 transits visible. With our detection efficiency independent of orbital period, we can examine the recovery rate as a function of the relative transit depth for the synthetic light curves. The transit depth is the true observable in the light curve, depending on both the planet radius and radius of the host star. Figure \ref{fig:sims_tdepth} presents the recovery rate of synthetic light curves as a function of the relative transit depth of the injected transits binned with a bin size of 3x$10^{-4}$ for transits with depths less than 0.005. We find as expected, detection efficiency drops at small transit depths. Beyond a relative depth of 0.0013 (relative transit depth of a $\sim$4 R$_{\oplus}$ planet around 1R$_{\odot}$ star), nearly all the simulated light curves are recovered. 3326 (97$\%$) of the  3417 synthetic light curves generated with transit depths larger the 0.0013 pass the candidate selection process and Round 2 review. 
 
\subsection{Comparison with the \emph{Kepler} Planet Candidates}
\label{comparekep}
A natural comparison to make is to the known list of KOIs identified by the \emph{Kepler} team's detection scheme. Our candidate detection scheme was not  tested or tuned to specifically identify known \emph{Kepler} planet candidates. Thus we can use the population of known KOIs to serve as an additional check of the Planet Hunters detection sensitivity.  We divide the KOIs into the same 7 radii bins ranging in 2-15  R$_{\oplus}$ and  6 orbital periods bins between 0.5 and 15 days as used to generate the synthetic light curves.  Of the 1235 KOIs identified in the first two quarters of \emph{Kepler} observations, 408 planet candidates are associated with multi-transit systems by \citep{2011ApJ...736...19B}.  Very few of the multiplanet systems satisfy our criteria  where we can be certain a single set of  transits in Q1 were identifiable to a Planet Hunters classifier, therefore we  restrict our analysis to only those currently identified to reside in single planet systems by \cite{2011ApJS..197....8L}. 330 KOIs fit our criteria and are used in our analysis presented here. We note that KOI 01.01 was missing from our Q1 sample of light curves and was not examined by Planet Hunters classifiers. 

Figure \ref{fig:kep} plots the Planet Hunters efficiency for the \emph{Kepler} sample of short period planets. Error bars are taken as the Poissonian 68$\%$ uncertainty (as prescribed by \cite{1991ApJ...374..344K} ) for the value in each radii/period bin. The  number of planet candidates and the number recovered by Planet Hunters is reported in Table \ref{tab:Keplerrecovery}. The KOI was deemed detected if it passed the Round 2 review process.  \cite{2012MNRAS.419.2900F} had visually inspected the classifications from the 306 \emph{Kepler} planet candidates announced by \cite{2010ApJ...713L.126B} finding that 2/3rds of the 1371 transits with planet radii between 1 and 10  R$_{\oplus}$ were flagged by Planet Hunters users with only 10$\%$ of transit boxes marking spurious detections. From the past visual inspection and low spurious detection rate, we are therefore confident that at least one of the appropriate KOI transits in each of the Planet Hunters candidate KOI light curves was identified. We note that  majority of  our comparison sample of KOIs are lost due to the transit score threshold cut with Round 2 review eliminating only 21 light curves and  the pulsator variability removing only 2 KOIs.

Although the error bars are large, we find that above 4  R$_{\oplus}$, Planet Hunters is  $>$ 80$\%$ efficient at identifying transits. We also observe the same trend  observed in the simulations.  Less than 4  R$_{\oplus}$ the detection efficiency drops with a 20$\%$ reported efficiency for the 2-3 R$_{\oplus}$ KOIs. Beyond 4  R$_{\oplus}$, the KOI recovery rates are in agreement with the detection efficiency measured for the injected transits. Additionally we see the same result we reported for the simulated transits, the detection efficiency is relatively insensitive to the number of transits present in the light curve. The largest discrepancy occurs in the 2-3  R$_{\oplus}$ bin, where only 20$\%$ of the \emph{Kepler} planet candidates are recovered. 40$\%$ of the simulated 2-3  R$_{\oplus}$  light curves were identified by Planet Hunters classifiers. 
  
  \subsection{2-3  R$_{\oplus}$ Detection Efficiency}

To further explore this discrepancy between the KOI recovery rate and the synthetic detection efficiency in the 2-3 R$_{\oplus}$ bin, we examine the detection efficiency as a function of relative transit depth for the 144 \emph{Kepler} and 1799 synthetic 2-3  R$_{\oplus}$ light curves. We expect that if the classifiers are able to detect a transit of a given depth from the Planet Hunters simulations, they should equally be able to detect a KOI of similar transit depth. The detection efficiency  for the 2-3 R$_{\oplus}$ simulated light curves as a function of relative transit depth is plotted as Figure \ref{fig:depth23}. As expected if the Planet Hunters classifiers are marking transits and not spurious features in the light curves, detection efficiency decreases with decreasing transit depth. Higher efficiencies are seen for transits with larger depth for both the simulations and the KOI distribution. 
 
 We test  if there are any significant differences between the detection sensitivity for the synthetic distribution and that obtained from the \emph{Kepler} planet candidates. We  bin the 2-3  R$_{\oplus}$  \emph{Kepler} KOI and synthetic samples into transit depth bins of 1.5x$10^{-4}$ in size.  We generate 100,000 `synthetic KOI' distributions by randomly selecting from the larger synthetic light curve population, the same number of simulations as there are real 2-3 R$_{\oplus}$ \emph{Kepler} KOIs in each transit depth bin. For each `synthetic KOI' sample generated, we determine which of the synthetic light curves were correctly identified as having transits by Planet Hunters. We then total the number of simulations recovered by Planet Hunters in each bin for the sample and compare with a chi-squared test to the binned histogram of Planet Hunters recovered real  2-3 R$_{\oplus}$  \emph{Kepler} KOIs.  We find there is no significant difference between the synthetic distribution and that obtained for the \emph{Kepler} planet candidates. Of the 100,000 `synthetic KOI' samples, 81.9$\%$ percent can be rejected as being drawn from the same population as the recovered KOIs at a significance greater than 68$\%$,  with only  9.6$\%$ rejected at the 95$\%$ confidence level. Therefore at the two-sigma level, the recovery rates for the  \emph{Kepler} KOIs and synthetic transits as a function of relative transit depth are consistent. 
  
 Nonetheless, when we plot the distribution of detected simulations and \emph{Kepler} candidates as a function of radius, we do see significant differences.  We use the Kuiper variant of the  Kolmogorov-Smirnov (KS) test, as detailed in \cite{1992nrca.book.....P}, to compare the  2-3  R$_{\oplus}$  KOI transit depth distribution to that of the synthetic population. The significance of the computed D statistic is found by performing  10,000 realizations of the KS test by randomly selecting the  number of real KOIs from the simulated transit depth  distribution and comparing to the full simulated sample and computing the fraction of instances where the computed D statistic was higher than the D statistic measured for the real KOI distribution. We find the hypothesis that both synthetic and KOI distributions are drawn from the same distribution can be rejected at the  99.98$\%$ confidence level. As sensitivity to transit depth is the same in both cases, the discrepancy in the recovery rate at the 2-3  R$_{\oplus}$ bin, can then  be attributed to differences in relative transit depth distribution between the two samples. 

The depth of the transit is the squared ratio between the  planet and stellar radii, and the simulations randomly sample the \emph{Kepler} target stellar population and have a uniform distribution of planet radii between 2 and 3 R$_{\oplus}$ bin. If there is a significant difference  in the distribution of stellar radii and planet radii for the 2-3 R$_{\oplus}$ KOIs this would explain the difference in the distributions. A 2-dimensional KS test \citep{1992nrca.book.....P,1983MNRAS.202..615P}  of the real and simulated population's planet radii and stellar radii distributions rules out at greater than the 95$\%$ confidence level that the two populations are compatible. The significance of the D statistic is estimated in the same manner for the 1-dimensional case, drawing 144 planets and their associated host star radii randomly from the synthetic population and comparing to the parent synthetic population 10,000 times and calculating the fraction of instances where the computed D statistic was higher than the D statistic measured for the real KOI distribution. Our simulation population fundamentally differs from the true underlying planet population, producing different transit depth distributions. 

The KOI planets are skewed towards smaller radii than the simulations. Additionally the KOI stellar radii are also slightly shifted towards lower radii compared to the simulations. These differences likely are representative of the true underlying planet population and formation environments. Though we note that any selection effect in \emph{Kepler} candidate selection, systematic errors in planet radii or stellar radii estimations would shift the planet radius distribution.

\section{Conclusions}

This paper presents results from our analysis of the first 33.5 days of \emph{Kepler} science data, focusing on the detection via transits of planets larger than or equal to 2  R$_{\oplus}$ with orbital periods shorter than 15 days. We use a set of   1,687,547 classifications provided by more than 24,000 citizen scientists using the Planet Hunters interface, we have performed  a systematic search for short period  planet candidates. After vetting 75 potential candidates identified via the Q1 classifications that were not identified as KOIs in  \cite{2011ApJ...736...19B} with an examination of  all currently available Kepler public data (Quarters 1-6), we find seven candidate light curves (KIC 551108, 6616218, 8240797, 9729691, 10905746 11017901, 11551692) not listed as planet candidates by  \cite{2011ApJ...736...19B} during the initial search of the first four months of Kepler observations but were identified in search of the first 16 months of \emph{Kepler} data \citep{2012arXiv1202.5852B} with an improved pipeline. KIC 11017901 was previously discovered and characterized by \citep{2012MNRAS.419.2900F} and the remaining six additional candidates have been identified  as containing  at least one $\ge$ 1.9 R$_{\oplus}$  planet candidate with transits repeating less than 15 days in later runs of the \emph{Kepler} pipeline searching  Quarters 1-6 \citep{2012arXiv1202.5852B, 2012arXiv1202.6007L}.  All seven candidates require additional follow-up observations to confirm their planetary nature. No additional candidates not listed as a KOI were found.
 
Using synthetic planet transits inserted into the data and the known  \emph{Kepler} sample of short period planet candidates, we have measured the detection efficiency of human classifiers to identify planet transits in the \emph{Kepler} Q1 light curves. Although performance drops rapidly for smaller radii,  $\ge$ 4 R$_{\oplus}$  Planet Hunters is $\ge$ 85$\%$ efficient  at identifying transit signals for planets with periods less than 15 days for the \emph{Kepler} sample of target stars. For 2-3 R$_{\oplus}$ planets,  the recovery rate for $<$ 15 day orbits drops to  40$\%$. Our high recovery rate of both $\ge$4 R$_{\oplus}$ simulations and KOIs and the lack of additional candidates not recovered by the improved \emph{Kepler} TPS and data validation routines and procedures suggests the \emph{Kepler} inventory of  $\ge$4 R$_{\oplus}$ short period planets is nearly complete.  

Although  our search was not specifically developed and optimized for identifying eclipsing binaries in the  Q1 \emph{Kepler} sample, we can comment on the \emph{Kepler} team's detection efficiency for the catalog of detached eclipsing binaries with orbital periods $<$ 15 days \citep{2011AJ....142..160S, 2011AJ....141...83P}.  Our detection scheme is highly sensitive to finding detached eclipsing binaries  where the separation distance between the two stars is much larger than their stellar radii, and the light curve has clearly defined and separated primary  transits and secondary eclipses. Only in our final stage of candidate selection, after Round 2 review, are the transit depths considered; any potential new eclipsing binaries would be found during the science team review.  Even for those light curves where the secondary eclipse is relatively obscured due to the vertical scaling needed to display the primary transit, the Planet Hunters volunteers would have marked the primary transit and in the final review stage the light would have been identified as an eclipsing binary. Most of the possible eclipsing binaries appear to be blends with a nearby eclipsing binary, leaving only a few possible new identifications. With 926 detached eclipsing binaries identified in the \cite{2011AJ....142..160S} catalog as having periods less than 15 days,  this would suggest the \emph{Kepler} eclipsing binary catalog has identified almost all eclipsing binaries with periods less than 15 days. 

Planet Hunters is a novel and complementary technique to traditional planet transit methods. The previous discovery of new planet candidates highlighted in \cite{2012arXiv1202.6007L} and  \cite{2012MNRAS.419.2900F}  and the high detection efficiency for  4 R$_{\oplus}$ and larger planets shows the efficacy of visual review of light curves for detecting large transiting planets. With a detection efficiency greater than 85$\%$ for 4R$_{\oplus}$ and larger planets orbiting on periods less than 15 days, it is likely Planet Hunters can easily detect these large planets even for single transits present in the 30-day light curve sections presented on the website. With the completion of Quarter 1, Quarters 2-5, each spanning approximately  90 days, have been added to Planet Hunters once publicly released, extending the observational baseline to $\sim$400 days, and classification of these light curves is currently underway. With the public release of Quarters 1-6, Planet Hunters will be able to examine the abundance of single transit events that do not repeat and would not be identified by TPS as well as search for additional transit events that may have been missed from longer period planets with repeating transits in Q1-Q6. However, transits due to planets significantly smaller than 2 R$_{\oplus}$ will likely remain difficult to detect;  particularly Earth-sized planets will be nearly impossible to identify by visual inspection with Planet Hunters due to the low signal to noise ratio of the individual transits. Improved user weighting schemes particularly promoting users who are more adept at identifying smaller radii planets may improve the detection efficiency at 2-3 R$_{\oplus}$  in the future, but the majority of  small rocky planet discoveries will likely remain in the realm of the traditional transit detection methods.  Human review may have a niche and be particularly valuable for systems which are unusual, such as circumbinary planets, or difficult because the presence of Transit Timing Variations (TTVs) prevents detection by automated routines which look for periodic repeats of transit features. \cite{2011MNRAS.417L..16G} highlight the difficulties such routines have with significant TTVs, and an application of the algorithm described here, which identifies individual transits rather than a periodic signal, to further releases of \emph{Kepler} data may be able to catch them.

\noindent{\it Acknowledgements} 

\noindent The data presented in this paper are the result of the efforts of the Planet Hunters volunteers, without whom this work would not have been possible. Their contributions are individually acknowledged at http://www.planethunters.org/authors. The authors thank the  Planet Hunters volunteers who participated in our Round 2 Review. They are individually recognized at \\ http://www.planethunters.org/Q1round2review.
\\
\\
MES is supported by a National Science Foundation Astronomy and Astrophysics Postdoctoral Fellowship under award AST-1003258. CJL acknowledges support from The Leverhulme Trust. DAF acknowledges funding support from Yale University and support from the NASA Supplemental Outreach Award, 10-OUTRCH.210-0001. KS acknowledges support from a NASA Einstein Postdoctoral Fellowship grant number PF9-00069, issued by the Chandra X-ray Observatory Center, which is operated by the Smithsonian Astrophysical Observatory for and on behalf of NASA under contract NAS8-03060. The Planet Hunters ÕTalkÕ discussion tool  was developed at the Adler Planetarium with support from the National Science Foundation CDI grant: DRL-0941610.  
\\
\\
We thank Tom Barclay and Darin Ragozzine for insightful and detailed manuscript comments and suggestions. We also thank Andrej Prsa for a thoughtful discussion of \emph{Kepler} eclipsing binaries. This paper includes data collected by the \emph{Kepler} spacecraft, and we gratefully acknowledge the entire \emph{Kepler} mission team's efforts in obtaining and providing the light curves used in this analysis. Funding for the \emph{Kepler} mission is provided by the NASA Science Mission directorate. The publicly released \emph{Kepler} light curves were obtained from the Multimission Archive at the Space Telescope Science Institute (MAST). STScI is operated by the Association of Universities for Research in Astronomy, Inc., under NASA contract NAS5-26555. Support for MAST for non-HST data is provided by the NASA Office of Space Science via grant NNX09AF08G and by other grants and contracts. This research has made use of the NASA Exoplanet Archive, which is operated by the California Institute of Technology, under contract with the National Aeronautics and Space Administration under the Exoplanet Exploration Program. The work presented here makes use of the PyKe software developed by Martin Still and Tom Barclay. 

\noindent {\it Facilities:} \facility{Kepler}

\appendix

\section{Appendix: Potential Biases in Candidate Selection and Detection Efficiency}
\label{ref:biases}
It is important to consider the potential biases in our analysis and determine how they may impact our candidate selection and user behavior. In this section, we discuss the effect of the interface upgrades and modifications on user classifications. Additionally we explore the potential biases that may effect the calculated synthetic detection efficiency.

The entirety of the simulations and a subset of the real Q1 light curves, have been screened, since  2011 Feb 1, after the interface upgrade enabling volunteers to draw transit boxes directly on the displayed light curve. With classifiers now drawing their own transit boxes rather than being given a predefined box size, the frequency of boxes drawn and the typical  transit box width and height likely changed. We do not place a stringent requirement on the number of transits identified  or the transit width in the selection criteria to identify candidates. Thus, there should be little effect on our results.  We only use the box width as a criteria during the weighting of the simulations, and since all the simulations were viewed in the updated interface there is no bias in the weighting scheme and the impact on candidate selection should be minimal. 

\begin{figure}
\epsscale{1}
\plotone{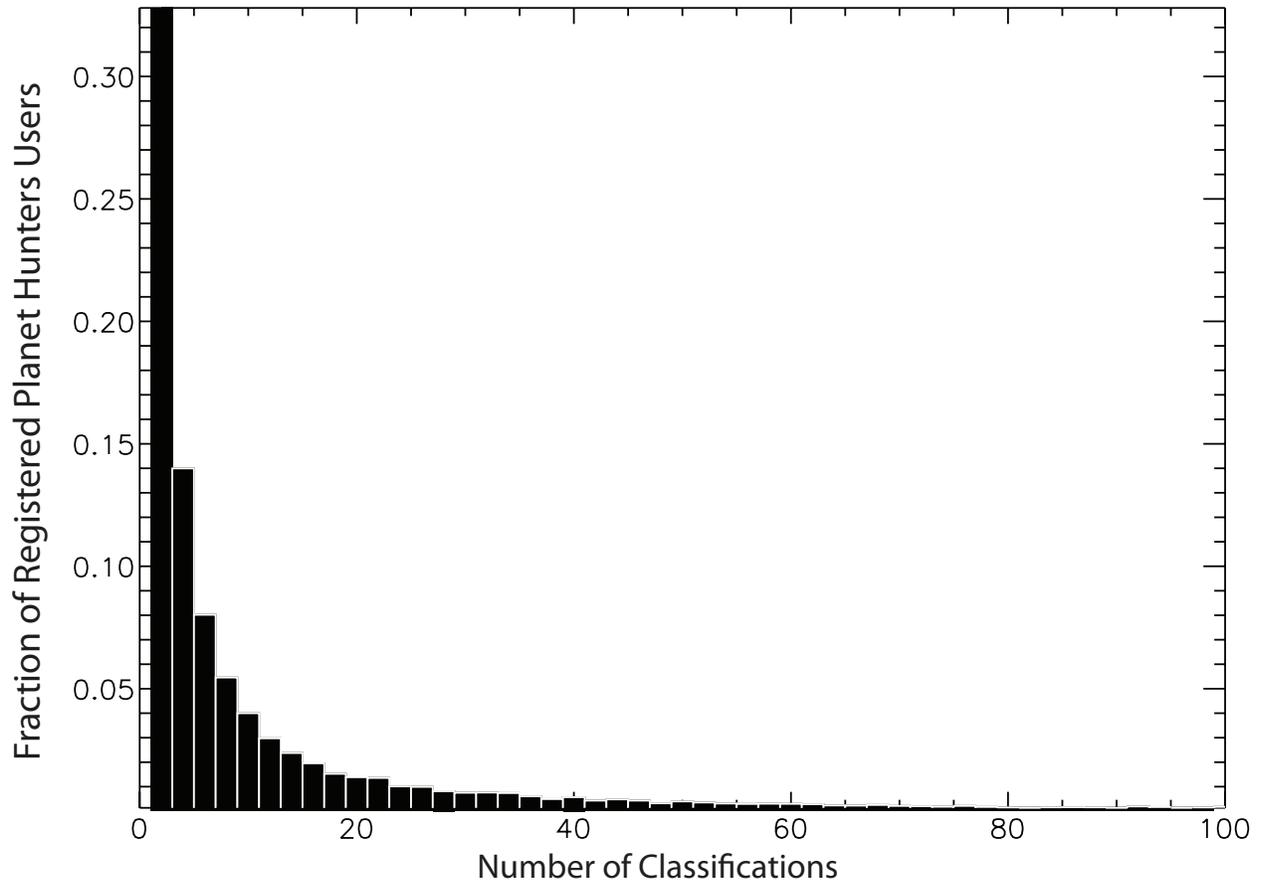}
\caption{Distribution of classifications used in this analysis per registered Planet Hunters user with a bin size of 2. The plotted distribution is truncated at 100 classifications for resolution. Only 10$\%$ of all registered volunteers make more than 100 classifications. }
\label{fig:fig4}
\end{figure}

\begin{figure}
\epsscale{1}
\plotone{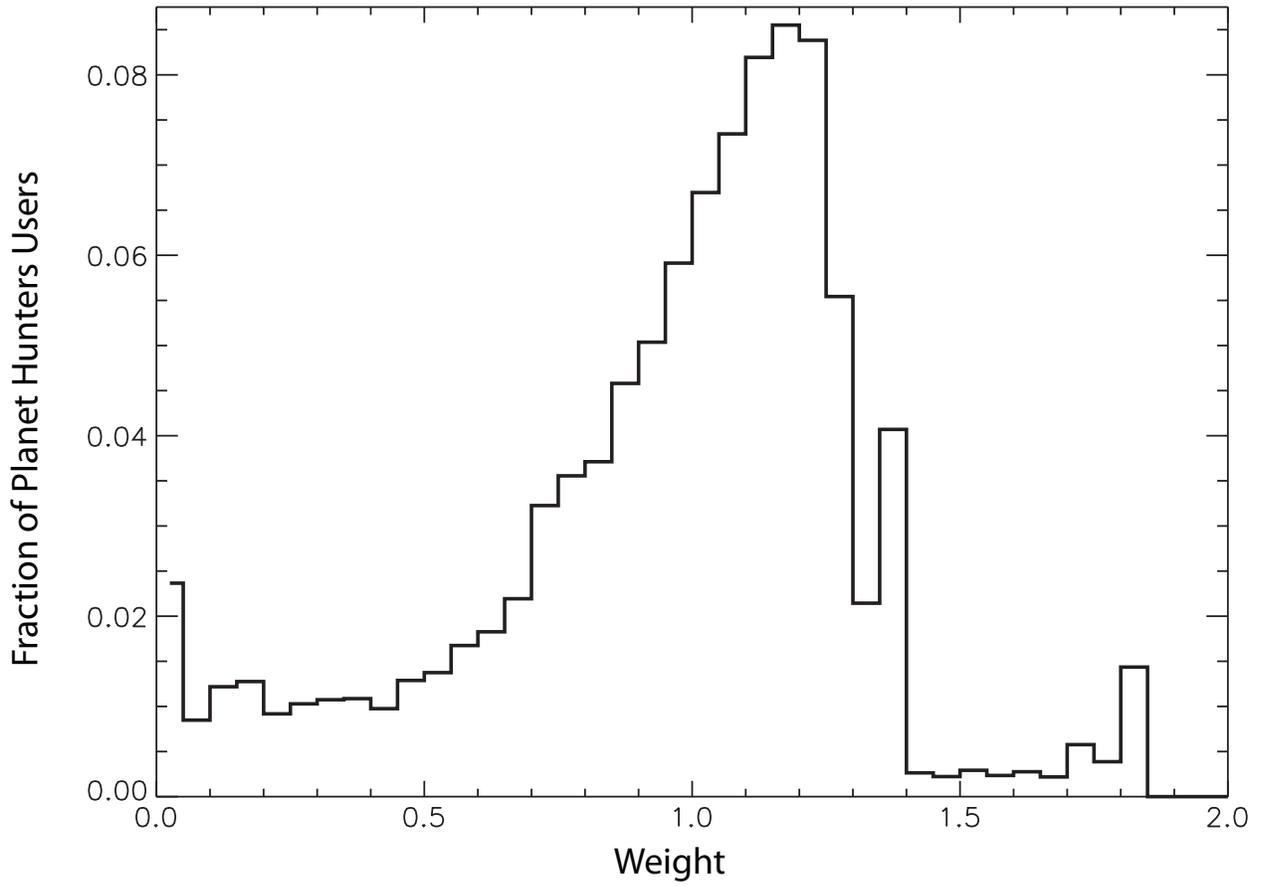}
\caption{Distribution of Q1 Planet Hunters user weights binned in 0.5 bins. }
\label{fig:fig5}
\end{figure}

\begin{figure}
\epsscale{0.7}
\plotone{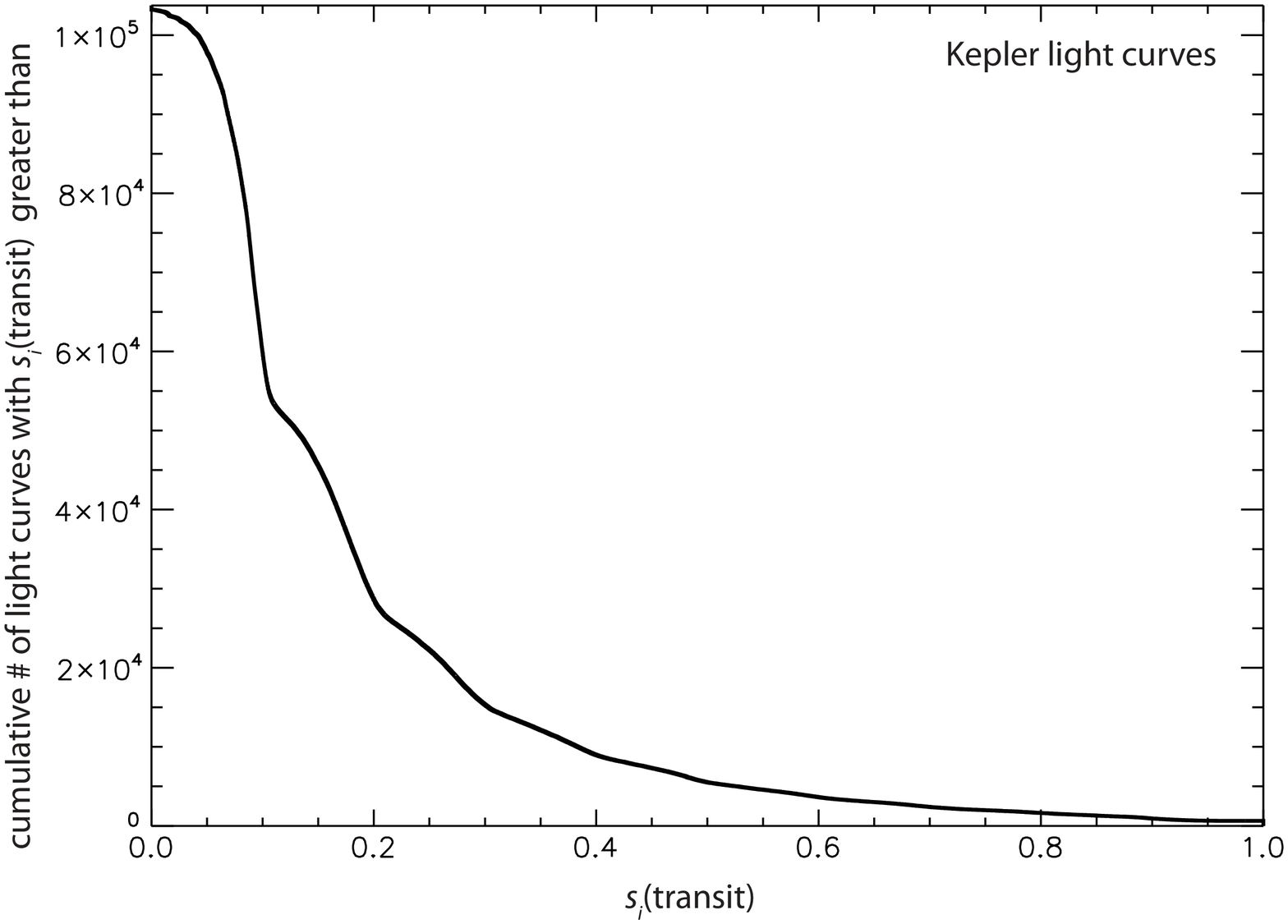}
\plotone{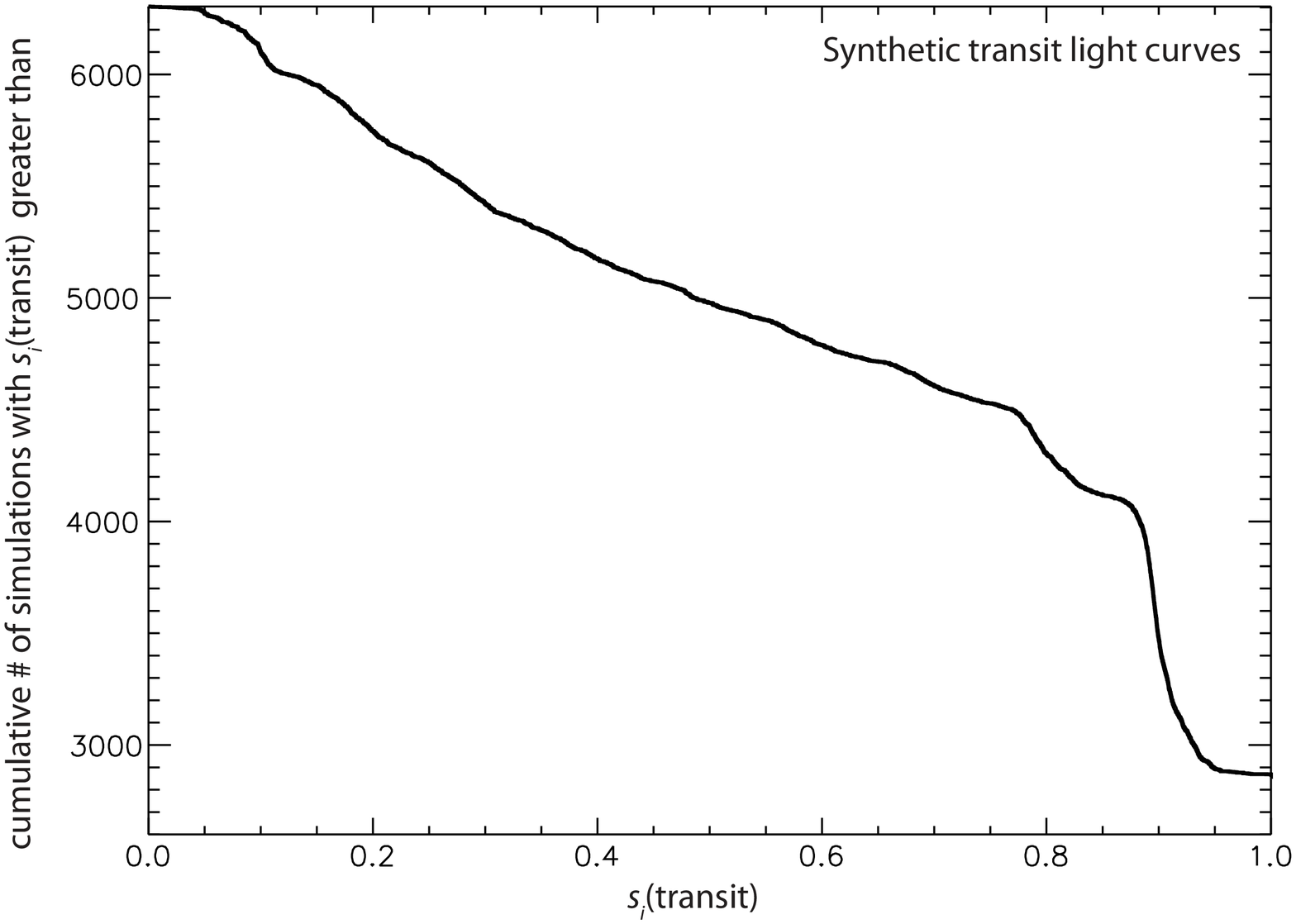}
\caption{ Cumulative distribution of transit scores for the Q1 \emph{Kepler} light curves (top) and simulations (bottom). }
\label{fig:fig6}
\end{figure}

\begin{figure}
\epsscale{1}
\plotone{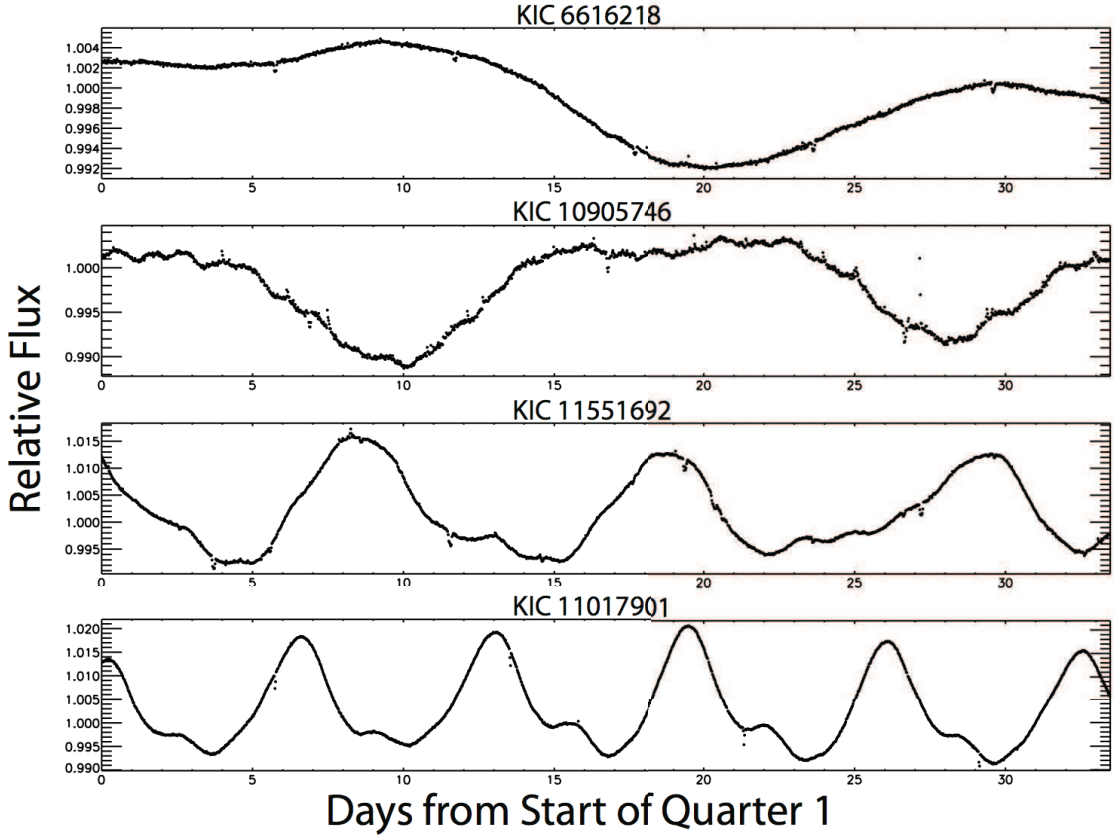}
\caption{ Short period candidate light curves  remaining after Round 2 Review and visual inspection. KIC 10905746 previously identified in Fischer et al 2012 and the remaining three candidates. All light curves were identified in later versions of the Kepler pipeline as KOIs in the latest KOI release \citep{2012arXiv1202.5852B}. KIC 11551692 and 6616218 have been subsequently identified by the \emph{Kepler} team as multiplanet systems with at least one $\ge$ 1.9$R_{\oplus}$  planet candidate orbiting the host star in less than 15 days (Batalha 2012 -personal communication, Batalha et al. 2012). The light curves have been normalized and a linear trend has been removed.  }
\label{fig:figcandidate_curves}
\end{figure}

\begin{figure}
\epsscale{1}
\plotone{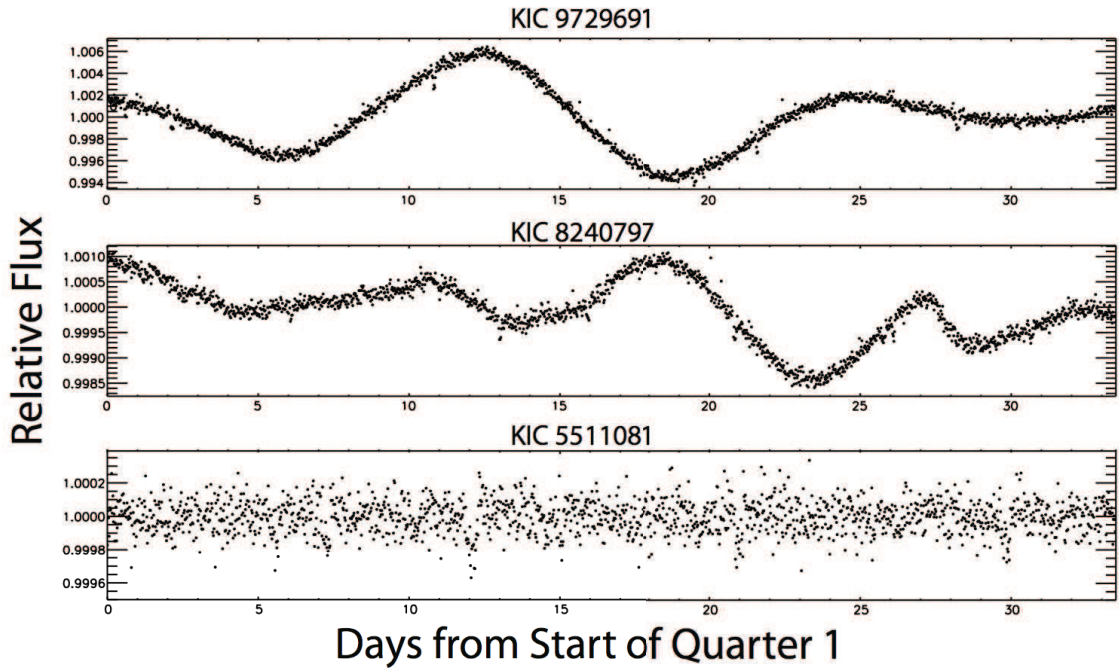}
\caption{ Short period candidate light curves remaining after Round 2 Review and visual inspection continued. All presented light curves were subsequently identified by the \emph{Kepler} team as multiplanet systems with at least one $\ge$ 1.9$R_{\oplus}$  planet candidate orbiting the host star in less than 15 days \citep{2012arXiv1202.5852B}. KIC 5511081 was also identified by Planet Hunters in a preliminary search of Q2 observations as reported by  \cite{2012arXiv1202.6007L}.  }
\label{fig:figcandidate_curves2}
\end{figure}

\begin{figure}
\epsscale{1}
\plotone{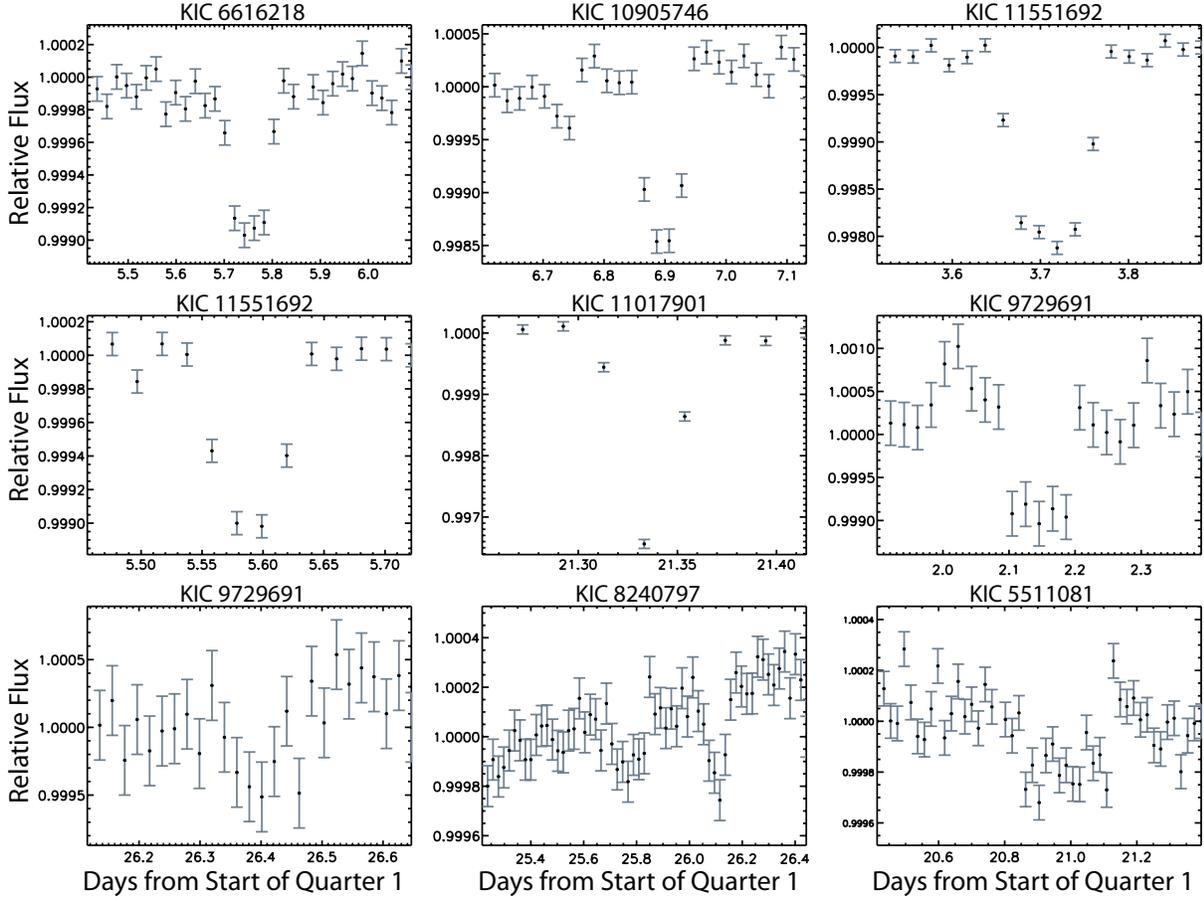}
\caption{ Zoom-in of selected transits  for each set of transit identified visible in short period candidate light curves remaining after Round 2 review and visual inspection. Visually the science team could identify two separate sets of repeating transits in the mutli-planet KIC  8240797, 9729691, and 11551692 based on the user drawn boxes We note that the snapshot of KIC 8240797 contains two independent transit events.  }
\label{fig:zoom}
\end{figure}

\begin{figure}
\epsscale{1}
\plotone{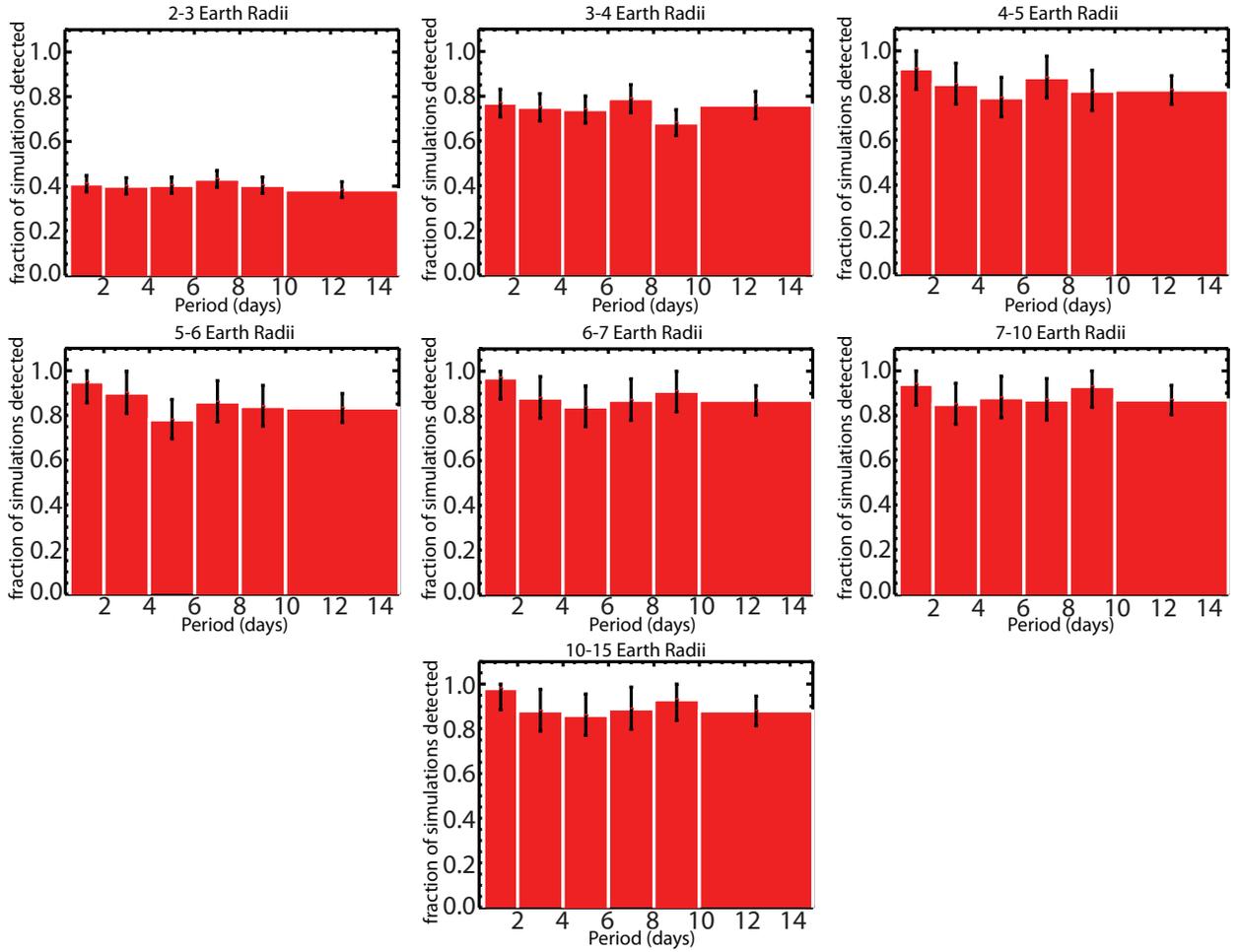}
\caption{ Efficiency recovery rate for simulated planet transits with orbital periods  between 0.5 and 15 days and radii between 2 and 15  R$_{\oplus}$. Error bars are taken as the Poissonian 68$\%$ uncertainty, as prescribed by \cite{1991ApJ...374..344K}, for the value in each radii/period bin.}
\label{fig:sims}
\end{figure}

\begin{figure}
\epsscale{1}
\plotone{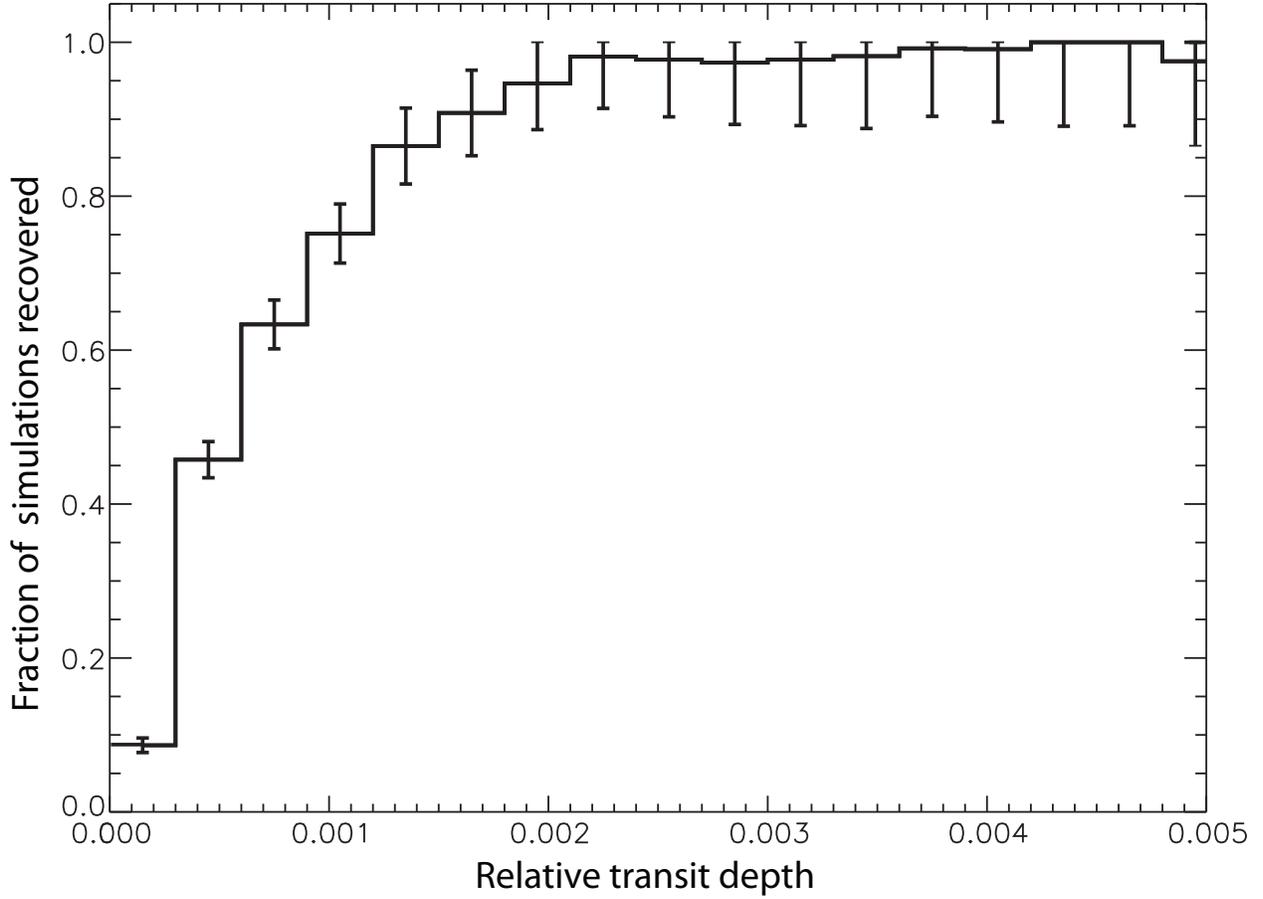}
\caption{  Recovery rate for simulated planet light curves for transits with depths less than 0.005, with orbital periods  between 0.5 and 15 days and radii between 2 and 15  R$_{\oplus}$,  as a function of relative transit depth binned with a bin size of 3x$10^{-4}$.  Error bars are taken as the Poissonian 68$\%$ uncertainty for the value in each bin, as prescribed by \cite{1991ApJ...374..344K}.}
\label{fig:sims_tdepth}
\end{figure}

\begin{figure}
\epsscale{1}
\plotone{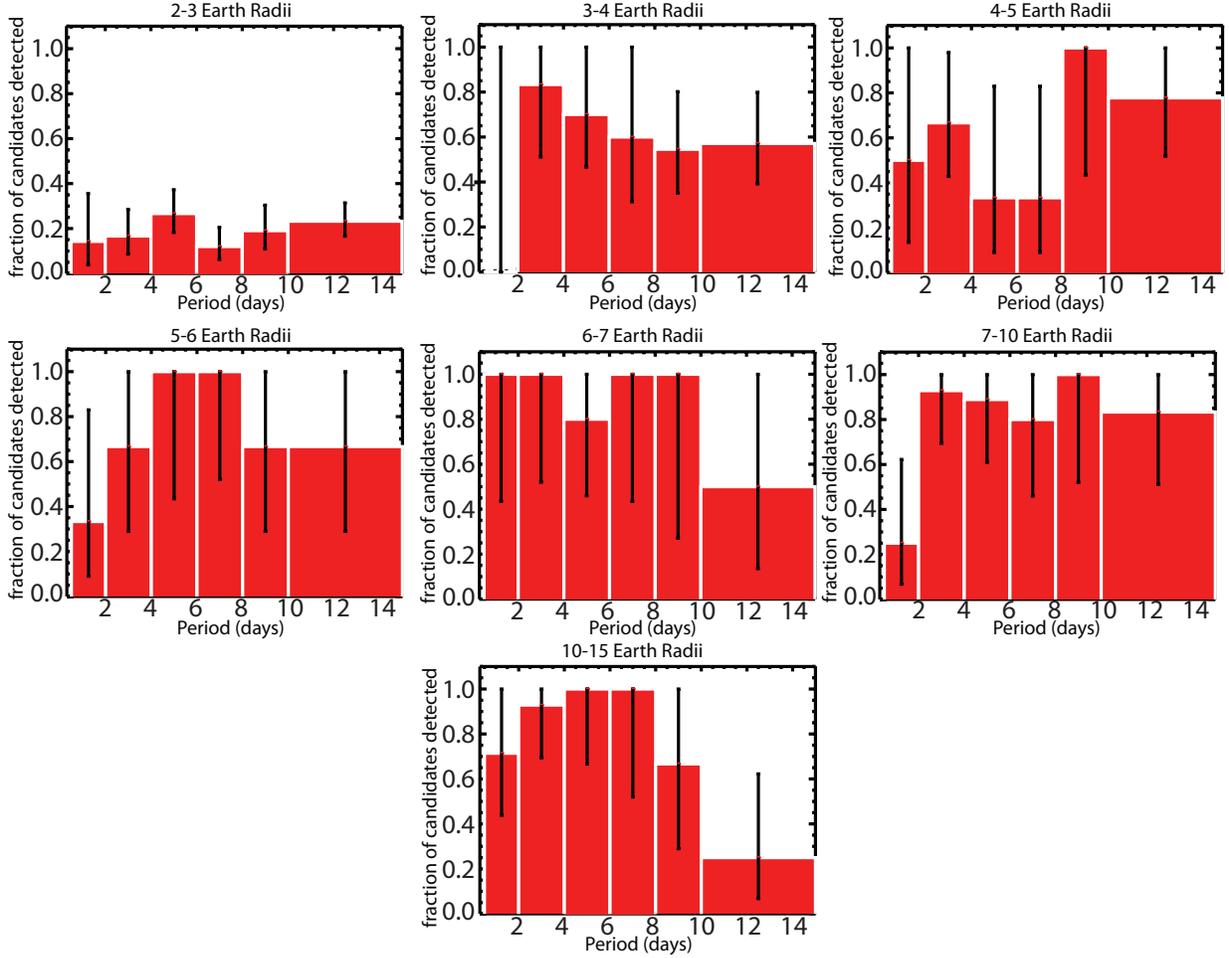}
\caption{ Recovery frequency of \emph{Kepler} planet candidates \citep{2011ApJ...736...19B} with orbital periods  between 0.5 and 15 days and radii between 2 and 15  R$_{\oplus}$ Error bars are taken as the Poissonian 68$\%$ uncertainty (as prescribed by \cite{1991ApJ...374..344K} ) for the value in each radii/period bin.}
\label{fig:kep}
\end{figure}

\begin{figure}
\epsscale{1}
\plotone{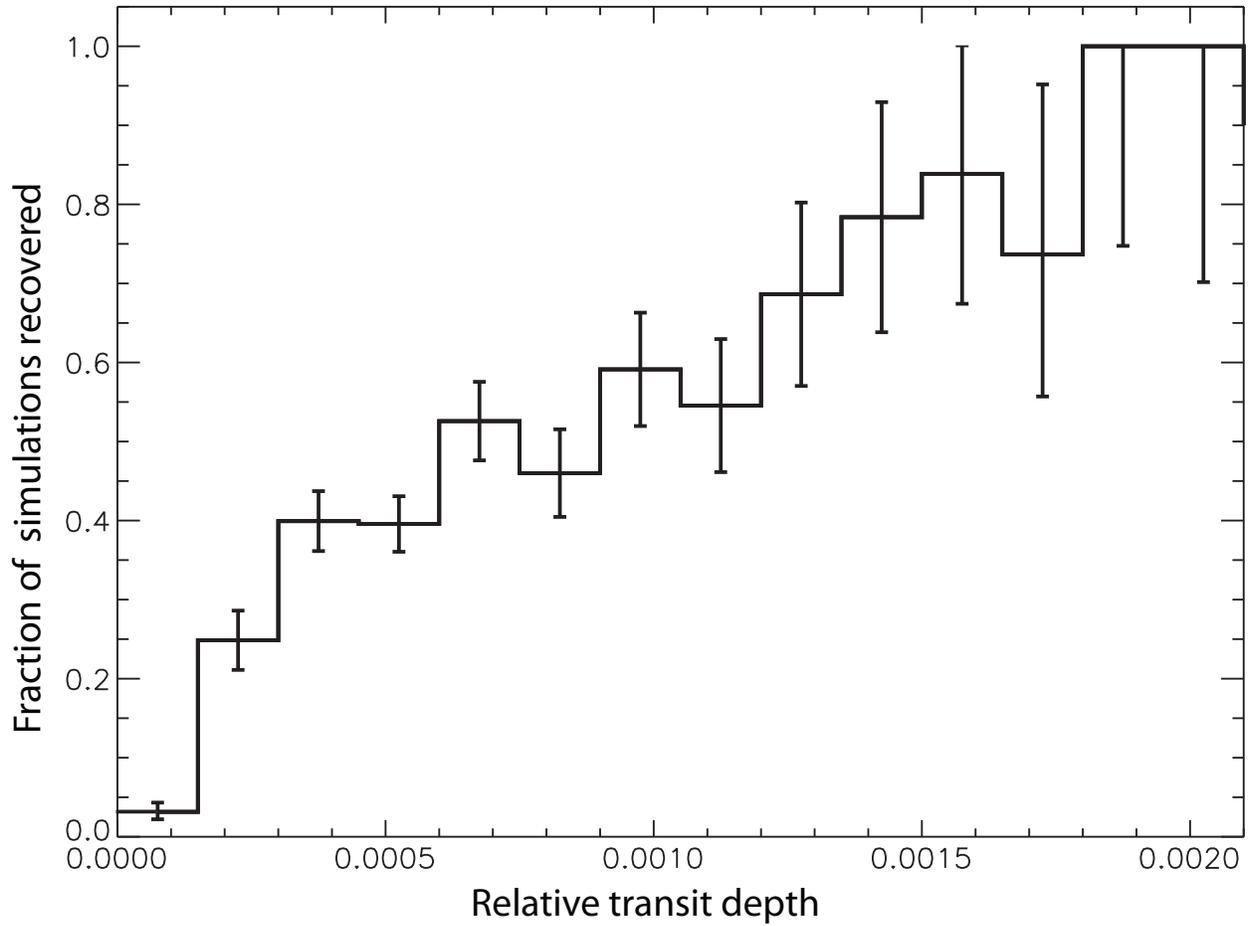}
\caption{ Recovery efficiency as a function of relative depth for  2-3  R$_{\oplus}$ simulations. For resolution the plot is truncated at .002. Error bars are taken as the Poissonian 68$\%$ uncertainty, as prescribed by \cite{1991ApJ...374..344K}, for the value in each bin.}
\label{fig:depth23}
\end{figure}

The subset of Q1 light curves and the simulations reviewed after the completion of the majority of Q1 data were interspersed with Q2 light curves which are scaled to a slightly different value and appear visibly different from the Q1 light curves. In particular, the Q2 light curves contain gaps due to missing data where as Q1 is continuous for a majority of the  \emph{Kepler} targets. This makes the Q1 data and simulations possibly recognizable such that when a user views a light curve without a gap he or she assumes the light curve is a simulation and is being tested therefore behaving differently than when the same user classifies real \emph{Kepler} light curves, i.e. the `Hawthorne effect' (Mayo 1933; Adair, Sharpe \& Huynh 1989). For example, the classifier may mark more transit boxes on features that the user is  less certain about because he or she wants to correctly identify the  `simulated' transits, and the simulations would then not reflect an accurate picture of the typical behavior of Planet Hunters volunteers. Most classifiers come to the Planet Hunters interface and classify a handful of light curves before leaving (see Figure \ref{fig:fig4}). Therefore most of the classifiers at any given time on the Planet Hunters classification interface will  be unaware of there being a difference between the two sets of light curves and what that significance is. Their responses will be the same whether  they are a classifying a synthetic light curve, Q1 light curve, or Q2 light curve. Although a small subset of users who have examined many tens of light curves may have noticed the pattern, these users remain a small fraction of the entire Q1 classifications. Only 13$\%$ of those users who classified a synthetic light curve classified more than 10 simulations. In addition the simulations and the remaining Q1 light curves were both shown at the same rate. 1 in 5 light curves served was either a Q1 or synthetic. Although the user might be able to identify the quarter of origin, they could not be certain if the light curve  presented was a simulation or real data. We believe this effect is negligible for the majority of our synthetic light curve classifications, and that they reflect the typical behavior of Planet Hunters classifiers.  

We also note that simulations were added to the Planet Hunters interface after 2011 May. The simulations were classified by a different set of users than were classifying the bulk of Q1. 34$\%$ of synthetic classifiers never examined a Q1 light curve. In order to estimate a detection efficiency,  we make the assumption that behavior between the two sets of classifiers is the same. Although there may be some evolution of the classifier population as a subset continue to classify curves and improve their ability to identify transits, the nature of the site, is such there is always new users who are classifying light curves. On average a visitor spends only 18 minutes on the Planet Hunters site. These new classifiers would be just as likely to review a simulation and have the same level of expertise as the original set of Q1 classifiers. Therefore, we feel they the synthetic classifiers serve as a representative sampling of the skill set and ability of the typical Planet Hunters classifiers, 154 of the 330 single planet KOIs with periods less than 15 days and  2-15  R$_{\oplus}$ were completed before the upgrade and were shown when the majority of the Q1 data was present in the classification interface. We can use  classifications as an independent test of our estimated synthetic detection efficiency. We find a similar recovery rate from these KOIs as for the injected transit light curves and full short period planet KOI sample. Therefore, we feel they the synthetic classifiers serve as a representative sampling of the skill set and ability of the typical Planet Hunters classifiers, and our detection efficiency calculated from the classifications of the synthetic light curves is appropriate for the entire Planet Hunters classifications.

\bibliographystyle{apj}

\begin{table}
\centering
\begin{tabular}{ c c c c c c c c c}
\hline
\hline
Planet Radii & \multicolumn{6}{c}{Orbital Period (days) } \\
\cline{2-7}
  ($R_{\oplus})$& $0.5\le P<2 $&  $2\le P<4$  &  $4 \le P<   6$&   $6 \le P< 8$ & $8 \le P<  10$ &  $10 \le P<  15 $\\

\hline

$2 \le  R < 3$ & 300 &  300 &  300 &  299 &  300 &  300  \\
$3  \le R <  4$  &  199 &   200 & 200  &  199   &  200&  200 \\
$4  \le  R <  5 $ &  100 &  100  &   100  &  100  &  100 &  200  \\
$5 \le R <  6$  & 100  &  100   &  100 &  100  &  100  &  198  \\
$6 \le R <  7$ &  100 &  100 &  100 &  100  &   100 &  200  \\
$7 \le R < 10$ & 100 &  100 & 100  &  100 &  100 &  200  \\
$10 \le R < 15$ &  100 & 100  & 100  & 99   &  100 &  200 \\

\hline
\hline
\end{tabular}
\caption{ Distribution of synthetic light curves generated}
\label{tab:sims}
\end{table}

\begin{table}
\centering
\begin{tabular}{ l | c |  c}
\hline
\hline
&   \multicolumn{2}{c}{ Remaining Light Curves } \\
\cline{2-3}
& Real & Simulations \\
  \hline

All &156,092 & 6494 \\

Transit Score Threshold &5508 & 4978  \\
Single Transit Removal  & 4938 &  4974 \\
Light Curve Variability Cut &  3404  &  4974 \\
Round 2 Review &  1220 & 4685 \\
Removing Known False Positives$^*$ & 379 & -- \\
Removing Known KOIs & 75&  --\\ 
Science Team Inspection & 7 & -- \\

   \hline
\hline
\end{tabular}
\caption{ Remaining light curves at each stage of the planet candidate selection process for both the Q1 \emph{Kepler} light curves and generated simulated light curves. $^*$False positives included eclipsing binaries and other false positives identified in \cite{2011ApJ...736...19B}, \cite{2011AJ....142..160S},\cite{2011AJ....141...83P} and \cite{2012MNRAS.419.2900F}. }
\label{tab:remaining}
\end{table}

\begin{table}
\centering
\begin{tabular}{ lccc }
\hline
\hline
Variability & $\#$ & fraction of  & fraction of   \\
class  & &   full population & variable population\\
\hline
unclassified & 849 & 0.005 & 0  \\
quiet &   102388 &  0.656  & 0 \\
 variable &  52855 & 0.339 & 1\\
  unclassified variable &  4260 &  0.027 &  0.081   \\
 regular variable &   7683 &   0.049  &  0.145 \\
 pulsating variable &  25820 &  0.165 &0.489  \\
 irregular variable &15092 &  0.098& 0.285 \\
 \hline
\hline
\end{tabular}

\caption{ Break down of light curve variability in the Q1 light curves based on Planet Hunters classifications }
\label{tab:variability}
\end{table}

\begin{table}
\centering
\begin{tabular}{ ccc }
\hline
\hline
&  $\le$16 mag   &  $\le$14 mag    \\
\hline
\cite{2011AJ....141..108C} & 35.2$\%$ & 60.8$\%$ \\
This work  &  32.0$\%$ &  49.6$\%$\\
 \hline
\hline
\\
\end{tabular}

\caption{ Variability frequencies  from \cite{2011AJ....141..108C}compared to that derived by Planet Hunters classifications. We quote the fraction of variable light curves from this work for the subset of light curves with viable variability assessments and had temperatures and radii estimated in the KIC \citep{2011AJ....142..112B} }
\label{tab:varcompare}
\end{table}

\begin{table}
\centering
\begin{tabular}{ lccc }
\hline
\hline
Variability & $\#$ & fraction of & fraction of  \\  
class  & &   full population & variable population\\
\hline
unclassified &  111& 0.022 & 0   \\
 quiet & 1835 &  0.372 & 0 \\
 variable & 2992 & 0.606 & 1 \\
  unclassified variable & 386  & 0.078 & 0.129  \\
 regular variable &   528 &  0.107 &  0.176 \\
 pulsating variable &  1776 & 0.360 &  0.594 \\
 irregular variable &302 &  0.061 & 0.101\\
 \hline
\hline
\\
\end{tabular}
\caption{ Break down of light curve variability in the Q1 planet candidate light curves based on Planet Hunters classifications }
\label{tab:varcand}
\end{table}

\begin{table}
\centering
\begin{tabular}{c c c c c c c c}

\hline
\hline
\multicolumn{8}{c}{ Single Event or Data Glitch or No Transits}\\
\hline

2161731 & 2557556 & 2569618 & 2992573 & 4261757 & 4846409 & 5437945\\
6117719 & 6186029 & 6603956 & 7023575 & 7175826 & 7457184 & 8297860 & 8584610 \\
8719324 & 8911948 & 9364562 & 9589420 & 10428163 & 11282332 & 11913545 & 12455830 \\
\hline
\\

\multicolumn{8}{c}{ Two or More Transit-like Features and No Consecutive Repeats in Q2$+$  Light Curves}\\
\hline
3337061 & 3937978 & 4552152 & 4739167 & 5212286 & 5561743 & 6032517 & 6219684  \\
7449421 & 7685675 & 7989590 & 8104436 & 8345153  & 8625821 & 9154469  &  9692612 \\
10467170 & 11516241\\
\hline
\\

\multicolumn{8}{c}{ Alternating Transit Depth Changes or Light Curve Contaminated by Eclipsing Binary}\\
\hline
2997178 & 3098197 & 3459199 & 3858879 & 3858917 & 4678875 & 5303346 & 5467124  \\
5471606 & 6032517 & 6045250 & 6182849 & 6543683 & 6612327 &  7877818 &  8095110  \\
 8104030 & 9529744 & 9851142 & 10031907 & 10092312 & 10275880 & 10294613 & 10735575  \\
11200767 & 11607176 & 12691412 \\
\hline
\\

\multicolumn{8}{c}{Remaining Short Period Planet Candidates }\\
\hline
5511081 & 6616218 & 8240797 & 9729691 & 10905746 & 11017901 & 11551692 \\ 
\hline
\hline 

\end{tabular}
\caption{Outcome of  Science Team Review- KIC 10905746  was previously identified by  \cite{2012MNRAS.419.2900F}. KIC 5511081, 6616218,  8240797,  9729691, 10905746, 11017901,11551692  have been identified in upgrades to the Kepler detection pipelines and  are to be included in the next  planet candidate release (Batalha 2012 personal communication, Batalha et al. 2012). KIC 5511081 was also identified by Planet Hunters in a preliminary search of Q2 observations as reported by  \cite{2012arXiv1202.6007L}. KIC 6186029 is a previously known RR Lyrae star \cite{2010MNRAS.409.1585B}.
}
\label{tab:round2}
\end{table}

\begin{table}
\centering
\begin{tabular}{ c c c c c c c c c}
\hline
\hline
Planet Radii & \multicolumn{6}{c}{Orbital Period (days) } \\
\cline{2-7}
  ($R_{\oplus})$& $0.5\le P<2 $&  $2\le P<4$  &  $4 \le P<   6$&   $6 \le P< 8$ & $8 \le P<  10$ &  $10 \le P<  15 $\\
   \hline

$2 \le  R < 3$ & 123/300 &  120/300 &  121/300 &  129/299 &  121/300 &  115/300  \\
$3  \le R <  4$  &  153/199 &   150/200 & 148/200  &  157/199   &  136/200&  152/200 \\
$4  \le  R <  5 $ &  92/100 &  85/100  &   79/100  &  88/100  &  82/100 &  165/200  \\
$5 \le R <  6$  & 95/100  &  90/100   &   78/100 &  86/100  &  84/100  &  165/198  \\
$6 \le R <  7$ &  97/100 &   99/100 &   84/100 &  87/100  &   91/100 &  174/200  \\
$7 \le R < 10$ &  94/100 &  85/100 &  88/100  &   87/100 &   93/100 &  174/200  \\
$10 \le R < 15$ &  98/100 &  88/100  &  86/100  &  88/99   &  93/100 &  176/200 \\
\hline
\hline
\end{tabular}
\caption{ Recovery frequency of  simulated light curves}
\label{tab:simsrecovery}
\end{table}

\begin{table}
\centering
\begin{tabular}{ c c c c c c c c c}
\hline
\hline
Planet Radii & \multicolumn{6}{c}{Orbital Period (days) } \\
\cline{2-7}
  ($R_{\oplus})$& $0.5\le P<2 $&  $2\le P<4$  &  $4 \le P<   6$&   $6 \le P< 8$ & $8 \le P<  10$ &  $10 \le P<  15 $\\
   \hline

$2 \le  R < 3$ & 1/7 &  3/18 &   8/30 &   3/25 &  4/21 &  10/43  \\
$3  \le R <  4$  &  0/ 1 &   5/6 & 7/10  &  3/5   &  6/11 & 8/14  \\
$4  \le  R <  5 $ &  1/2 &  6/9  &   1/3  &  1/3  &  2/2 &  7/9  \\
$5 \le R <  6$  &  1/3   &  2/3   &   2/2 &  3/3  &  2/3 &  2/3  \\
$6 \le R <  7$ & 2/2 &   3/3 &   4/5 &  2/2 &   1/1 &  1/2  \\
$7 \le R < 10$ &  1/4 &  13/14 &  8/9  &   4/5&   3/3 &  5/6  \\
$10 \le R < 15$ &  5/7 &  13/14  &  7/7  &  3/3  &  2/3 &  1/4 \\
\hline
\hline
\end{tabular}
\caption{ Recovery frequency of \emph{Kepler} planet candidates \citep{2011ApJ...736...19B} with periods  between 0.5 and 15 days and radii between 2 and 15  R$_{\oplus}$}
\label{tab:Keplerrecovery}
\end{table}

\FloatBarrier

\end{document}